# STL4IoT: A Statechart Template Library for IoT System Design

**Clyde Rempillo[1] and Sadaf Mustafiz[2]**

**Abstract**

The engineering of IoT systems brings about various challenges due to the inherent complexities associated with such heterogeneous systems. In this paper, we propose a library of statechart templates, STL4IoT, for designing complex IoT systems. We have developed atomic statechart components modelling the heterogeneous aspects of IoT systems including sensors, actuators, physical entities, network, and controller. Base system units for smart systems have also been designed. A component for calculating power usage is available in the library. Additionally, a smart hub template that controls interactions among multiple IoT systems and manages power consumption has also been proposed. The templates aim to facilitate the modelling and simulation of IoT systems. Our work is demonstrated with a smart home system consisting of a smart hub of lights, a smart microwave, a smart TV, and a smart fire alarm system. We have created a multi statechart with itemis CREATE based on the proposed templates and components. A smart home simulator has been developed by generating controller code from the statechart and integrating it with a user interface.

## Introduction

The Internet of Things (IoT) paradigm has brought about new challenges in software and systems engineering due to the complex, heterogeneous systems that are required to cooperate reliably in an uncertain and ever-changing environment. IoT is revolutionizing application domains from consumer and commercial to industrial and infrastructure (1). Smart home systems have gained immense popularity with applications such as, smart lights, smart appliances (smart fridge, smart washing machine, smart microwave, etc.), smart TV, and smart garage. Smart home automation aims to control the numerous systems used within a home as well as monitor energy consumption autonomously (2).

The different facets of IoT - software, hardware, network, and environment - need to be addressed in the design of such complex systems. The cyber-physical and heterogeneous nature of things and the wide array of possible applications make simulation essential for the design and deployment of smart services ((3), (4)). We also need to design our system using the most appropriate modelling languages at the most appropriate levels of abstraction.

Statecharts (5) are commonly used for behavioural modelling, simulation, and code synthesis. The behaviour of each smart system along with the coordination of the different systems can be modelled and simulated as a multi state machine system. The complex design of such systems includes various generic components modelling the hardware, software, network, and environment, and the interplay among them. Having templates that provide guidelines and gives a head start in modelling IoT/CPS system behaviour would be quite beneficial for designers. This would also be helpful for modellers who do not have deep knowledge on the heterogeneous components of such systems.

Previously, we have proposed a statechart template library to model the behaviour of IoT systems. The templates come with a library of atomic components to model various aspects of IoT systems, namely sensors, actuators, controller, network, as well as an additional component to be used for modelling power consumption. The goal of the template is to facilitate the process of modelling complex IoT systems by assisting designers with building multi state machines to detail the system behaviour and to reduce development time and effort. As an illustrative example, we developed a training simulator for a smart home system using our library of templates.

This paper is an extension of (6). It introduces new features in the template library along with an extended and more complex smart application. We have extended the template library with a template for physical entities. The illustrative example of the smart home application has been extended with a smart hub of lights. We have expanded the description of each system within the smart home with requirements models and added an architecture diagram of our smart home system. Models of the smart microwave and smart TV systems are also provided. We have further validated the usefulness and reusability of our hub template with a smart lights hub to demonstrate that the template

[1]Toronto Metropolitan University, ON, Canada
[2]Toronto Metropolitan University, ON, Canada

**Corresponding author:**
Sadaf Mustafiz, Department of Computer Science, Toronto Metropolitan University, 245 Church Street, Toronto, ON, Canada.
Email: sadaf.mustafiz@torontomu.ca

can be used to model hub of hubs. Moreover, we provide more details on the templates and process, and also expanded the related work section with a detailed comparison table. A demo video covering the new features and the extended example has been made available.

This paper is structured as follows: Section 2 provides essential background information. Section 2 outlines the running example of the smart home application. Section 4 presents the statechart templates for IoT. Section 10 demonstrates the application of our templates to the smart home system. Finally, Section 14 discusses related work and Section 17 concludes the paper and mentions some future work.

## Background

Our work in this paper is based on the statecharts language. In this section, we provide a brief background on statecharts and our target application domain, IoT.

### *Statecharts*

The Statecharts formalism is an extension of Deterministic Finite State Automata with hierarchy, orthogonality and broadcast communication (5). It is a popular formalism for the modelling of the behaviour of reactive systems. It has an intuitive yet rigourously defined visual notation and semantics. It is the basis for documentation, analysis, simulation, and code synthesis. A statechart model is usually described with the following basic elements: states (basic, orthogonal, composite), transitions (event-based or time-based), enter/exit actions, history state, guards or conditions, and actions. Please refer to (5) for details. Many variants of Statecharts exist, including the one in the UML standard (7).

Itemis CREATE (8) (formerly called Yakindu) is a modular toolkit for modelling and simulating statecharts. It also supports generation of executable finite state machines (FSM). Itemis CREATE Statechart Tools is based on Harel's Statecharts and supports the Statemate semantics introduced by Harel. It also supports multi state machine modelling that enables complex behavioural models to be modularized by splitting the model into smaller state models. This allows for separation of concerns and for the models to be reused by embedding in other statecharts. Multiple instances of a given statechart can be created. A system can thus be modelled as multiple collaborating statecharts. Itemis CREATE statecharts hold structural properties in addition to the behavioural aspect. These properties include variables and events that define the interface of the statechart. CREATE provides code generation support for the Java, Python and C languages among others.

### *IoT Systems/Cyber-Physical Systems*

**IoT/Cyber-Physical Systems (CPS).** In this paper, we follow the unified definition of IoT/CPS proposed in (9): *Internet of Things and cyber-physical systems comprise interacting logical, physical, transducing, and human components engineered for function through integrated logic and physics.*

The *things* constituting an IoT system (or corresponding "physical" entities in a CPS) have common goals and need to interact and cooperate to fulfill the functionalities of a smart system. The need to communicate with and control physical devices assigns new characteristics to such systems. The many facets of IoT - software, hardware, network, and environment - have to be taken into consideration during system design. An IoT/CPS system has a logical state and a physical state. A change in physical state trigger sensors to produce values for the logical system, which may lead to changes in the logical state. This may in turn trigger actuation events that lead to a change in physical state.

*In this paper, although we refer to the target domain as IoT, our work applies to both IoT and smart CPS.*

## Running Example

In this section, we present our running example of a smart home application, which is a representative case for IoT systems (10; 11). Our smart home system is composed of a smart fire alarm system, a smart lights system, a smart TV system, and a smart microwave system. All systems are also integrated together into a central smart hub.

### *Smart Fire System*

The *Smart Fire Alarm System* continuously monitors the heat, smoke, and carbon monoxide levels within an environment. We simulate a home environment which is monitored by the *Smart Fire Alarm System*, and if any risk of danger is sensed by the sensors, the fire alarm is activated. The following describes the system behaviour.

- Smart Fire Alarm System must always start from a `safe` state.
- The `sensor` component continuously monitors the environment for any high risk or danger levels.
- While monitoring the smoke, heat, and carbon levels, the controller determines whether any of the levels are above threshold.
- If the threshold value is crossed, the system will enter warning mode.
- When the system is in warning mode, it transitions to the `warning` state and activates the `timer`.
- The system provides two warning states: `initial warning` and `final warning` with corresponding timers. Each timer state has their own timer values.
- If the the timer runs out for both warning states, then the system is transitioned to `danger` mode.

### *Smart Lights System*

The *Smart Light System* is an IoT system composed of sensors and actuators. An ultrasonic motion detector is used to detect activity within the light's vicinity. When it detects movements, the actuators come into action to turn on and manage the lights. The *light system* features an integrated `WiFi` component that simulates an internet gateway connection to the server, which allows the hub to connect to the system.

Moreover, to keep track of how much energy the system is using, we have added a `power` component that calculates the energy consumption in kilowatt-hours (kWh). This energy data can be sent to a hub for efficient control and optimization.

The following describes the behaviour of the system:

- Smart Lights System will always start from an `off` state.
- Using the `ultrasonic motion detector`, it will detect physical movements using the built-in transmitter and receiver.
- The motion detection triggers an actuating event to activate the light.
- While the lights are on, the system has a timeout caused by inactivity which is continuously monitored by the sensors. After this timer runs out, the lights will automatically be turned off.

## *Smart TV System*

The *Smart TV System* is a smart system that models a television system that has sensors and actuators. The system can either be controlled through the hub or manually through the TV unit. The system includes basic TV functionalities such as satellite channels, HDMI connection and more. The TV's `sensor` component is only triggered for inactivity purposes, to help prevent the problem of continuous TV usage during inactivity. For clarity, the `sensor` component will not send any trigger signal to the `controller`. After a certain number of seconds of inactivity, the system will prompt the user to check if the TV is still being used. After the user's confirmation, the system will stay active, however, if there is no response received after the `timeout` value, then the system will automatically consider this as an inactivity and then the `controller` will trigger an actuating event to turn off the TV. The following describes the behaviour of the smart TV system.

- The system is initially off and can be turned on or off manually directly at the unit or remotely through the hub.
- The motion sensors are turned on only when the system is on and will continue monitoring for activity.
- When the sensors cannot detect any physical activity in the nearby surroundings, it will start the timeout counter using the timer component within the system. After the timeout, the system will request for activity feedback from the user.
- If no response is received, then after the timeout value, the system will be turned off automatically. However, if the user responds, then the system will stay on and cancel the timeout counter.

## *Smart Microwave System*

This *Smart Microwave* emulates the functionality of a traditional microwave oven while incorporating smart features. In addition to fundamental operations like opening and closing, setting timers, and initiating or halting the microwave, this system introduces smart capabilities. It can detect the presence of food inside the microwave using sensors, conserves power during periods of inactivity, and establishes network connectivity via WiFi. The following list describes the behaviour of the smart microwave system.

- The smart microwave starts in the *idle* mode once switched on by the user.
- The system will start activity when the door is opened. It will return to *idle/standby* mode if there is no activity after the timeout using the timer component.
- The *sensor* component enables sensing of any presence of food to ensure that the microwave is not used when empty.
- The system also includes a heat sensor to monitor the internal temperature and prevent explosive accidents with the system. This allows the system to automatically turn the system off once it reaches the temperature threshold.

## *Smart Hub System*

The *Smart Hub System* is an IoT system that acts like a hub and integrates multiple smart systems and allows each one to communicate and be controlled through the hub. In our running example, the smart hub for the smart home is composed of these smart systems: Smart Fire Alarm, Smart Lights, Smart TV, and Smart Microwave.

Each system works autonomously and are interconnected through the hub, which acts as the central terminal that manages coordination of the systems. The hub device enables the user to switch on/off the smart systems, with the exception of the smart fire alarm, remotely.

If the Smart Fire Alarm System activates the fire alarm when fire is detected, then it will notify and alert the hub. Then, the Smart Hub will process this emergency procedure and deactivate all of the other connected systems and restrict them from being operated until the fire alarm is turned off and safety is confirmed.

Other hub features such as `Power Manager` is also implemented to manage the power consumption level of the home environment and ensure that the total power consumption level stays under threshold. If the total consumption crosses the threshold, the manager recognizes this and turns off the most consuming system until the total is under threshold again. It simulates a power threshold manager designed within the `Power Manager` as it turns off one or more of the smart devices when the threshold is reached or exceeded.

The smart hub services are summarized in Fig. 1 and the required behaviour is detailed below.

- The *Smart Hub System* enables the user to switch on/off the smart systems remotely including the *Smart TV System, Smart Light System* and the *Smart Microwave System.*
- Using the `Power Hub Manager`, the system manages the power consumption level of the home environment and ensures that the total power consumption level stays under threshold. If the total level exceeds, then the `Power Hub Manager` recognizes this and turns off the most consuming system until the total is under threshold again.
- When the *Smart Fire Alarm System* sends a critical message to the hub, the *hub* deactivates all of the other smart systems including the *Smart TV System*, *Smart Light System* and the *Smart Microwave System.* The *hub* restricts the user from using these system until the danger is handled and the fire alarm is turned off.

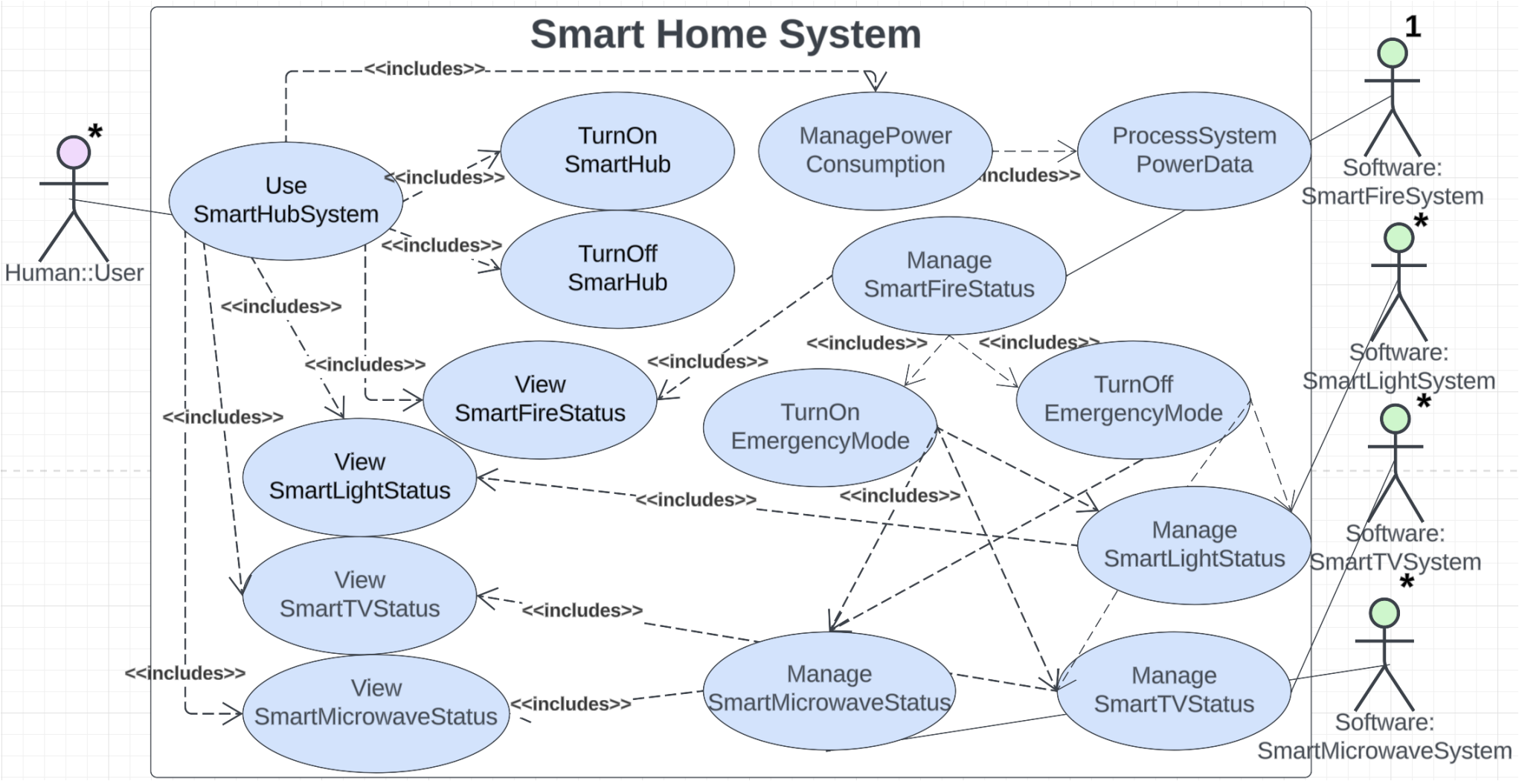

**Figure 1.** Smart hub system: use case diagram.

## *Smart Hub of Hubs*

The Smart Hub of Hubs (SHoH) is a central hub that serves as a unifying point for connecting and controlling multiple other specialized hubs in a network or smart home ecosystem. These specialized hubs are often designed for specific functions, such as lighting, security, entertainment, or heating and cooling. The central hub acts as a coordinator, allowing these different hubs to communicate with each other and be controlled through a single interface. Each specialized hub is responsible for controlling and managing a particular set of smart devices. For example, you might have a lighting hub that controls all your smart light bulbs, a security hub for cameras and sensors, and a media hub for your entertainment system. Moreover, there is the central hub, which connects to all these specialized hubs. It acts as a bridge, translator, and controller, allowing you to interact with and automate the various devices connected to the specialized hubs. Hence, the central hub enables integration between different devices and systems. As part of this running example, we use a hub of smart lights that is connected to the smart home system.

- The *SHoH* enables the user to switch on/off multiple smart lights remotely.
- Using the `Power Manager` in the hub of lights, the system manages the power consumption level of the home environment and ensures that the total power consumption level stays under threshold. If the total level exceeds, then the `Power Hub Manager` recognizes this and turns off the most consuming system until the total is under threshold again.
- When the *Smart Fire Alarm System* sends a critical message to the hub, the *central hub* deactivates all of the other smart systems including the *hub of lights*.

## *Smart Home Architecture*

The architecture diagram shown in Fig. 2 describes the overall structure of the framework and the interactions between each smart system and its' components. The model specifically describes the data flow between the *Smart Hub System* and the other connected smart systems. The `system` manager has direct connection with the smart systems via network connectivity. All of the smart systems send their system data to the hub's system manager, and receives data commands from the hub via system manager, as shown in Fig. 2. For example, when the *Smart Fire System* goes into `Emergency Mode` due to danger risk. It sends this data to the *Smart Hub System* through the `System Manager`. The hub will then use this data to enter `Emergency Mode` and then send a command to all smart systems to turn off until the danger is resolved.

## Statechart Template Library

In this section, we introduce, STL4IoT, a library of templates and components for modelling the behaviour of IoT systems. The hub and IoT templates are skeletons used for building statecharts for smart systems for the purpose of simulation. They can be easily modified or extended at any time according to the complexity of the system. The library can be utilized to model various smart IoT systems without having to build from scratch. This library contains a growing collection of components and base systems that can be considered as building blocks towards developing complete, sophisticated, and complex IoT systems. We have taken inspiration from the IoT literature, such as (12), (13) and (14), and designed reusable *atomic components* and *base system units* that is applicable for a wide range of IoT applications.

A component can be of two types: 1) atomic, and 2) base system unit. Using the IoT Template, we follow the structure within the template and import specific components to essentially build the system under study. For example, for the Smart Fire System, we used the IoT Template to design the complete system, however, we imported the atomic components from the library, namely the sensor, actuator and controller. We also used the base system unit for the Smart Fire System. Essentially, without the integration of the IoT template and the components, the base units are just individual units working autonomously. Moreover, we also developed a Smart Hub Template for IoT systems that plays the role of a coordinator or hub for multiple systems. The hub template can be extended by adding desired systems.

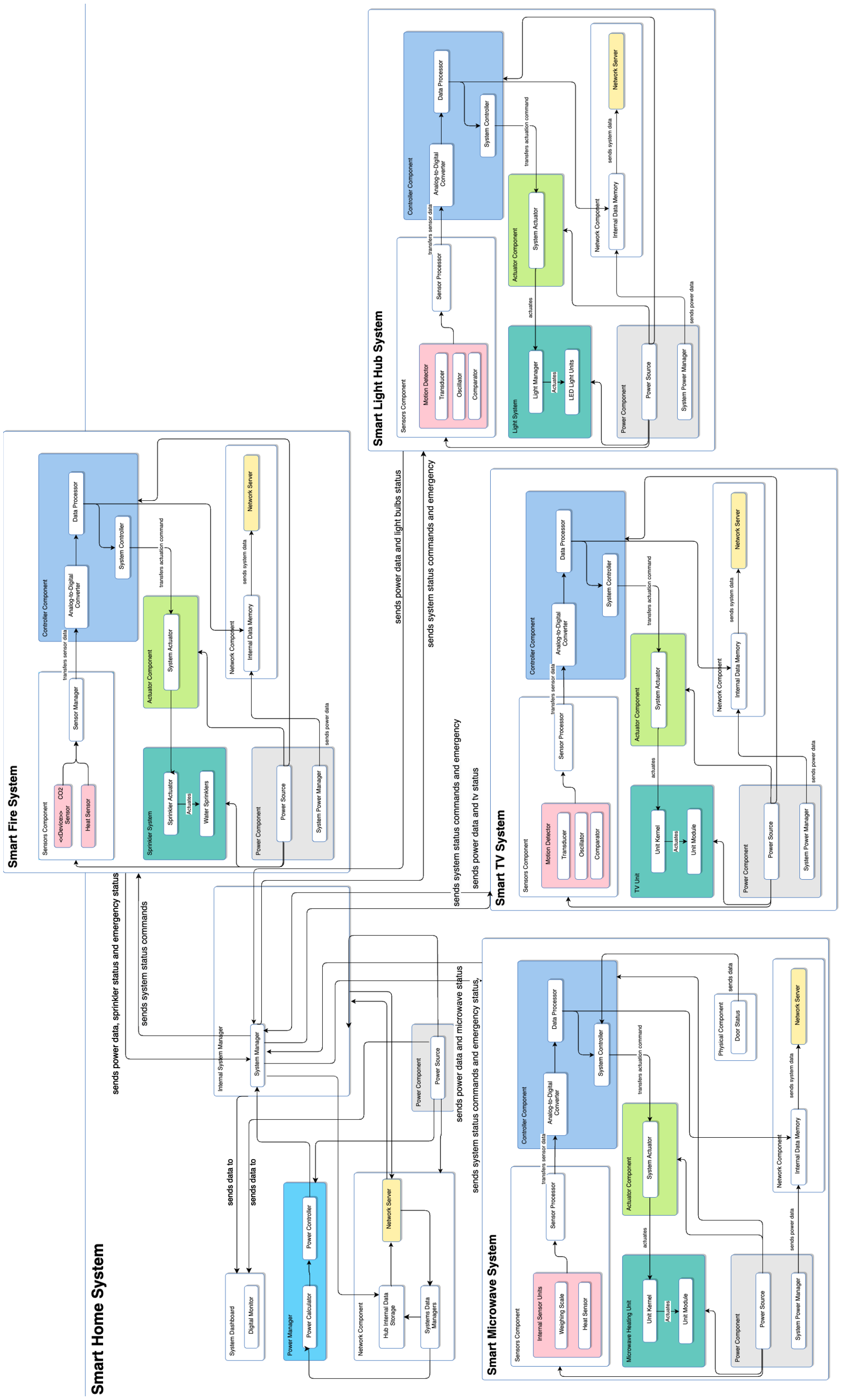

**Figure 2.** Smart home: architecture diagram.

The STL4IoT library along with the components and dependencies are shown in the concept map in Fig. 3.

The library of templates has been implemented with Itemis CREATE Statechart Tools (8) (see Fig. 4). CREATE provides support for simulation of smart system models constructed with our library and for the generation of statechart code from these models. A video demonstrating the template

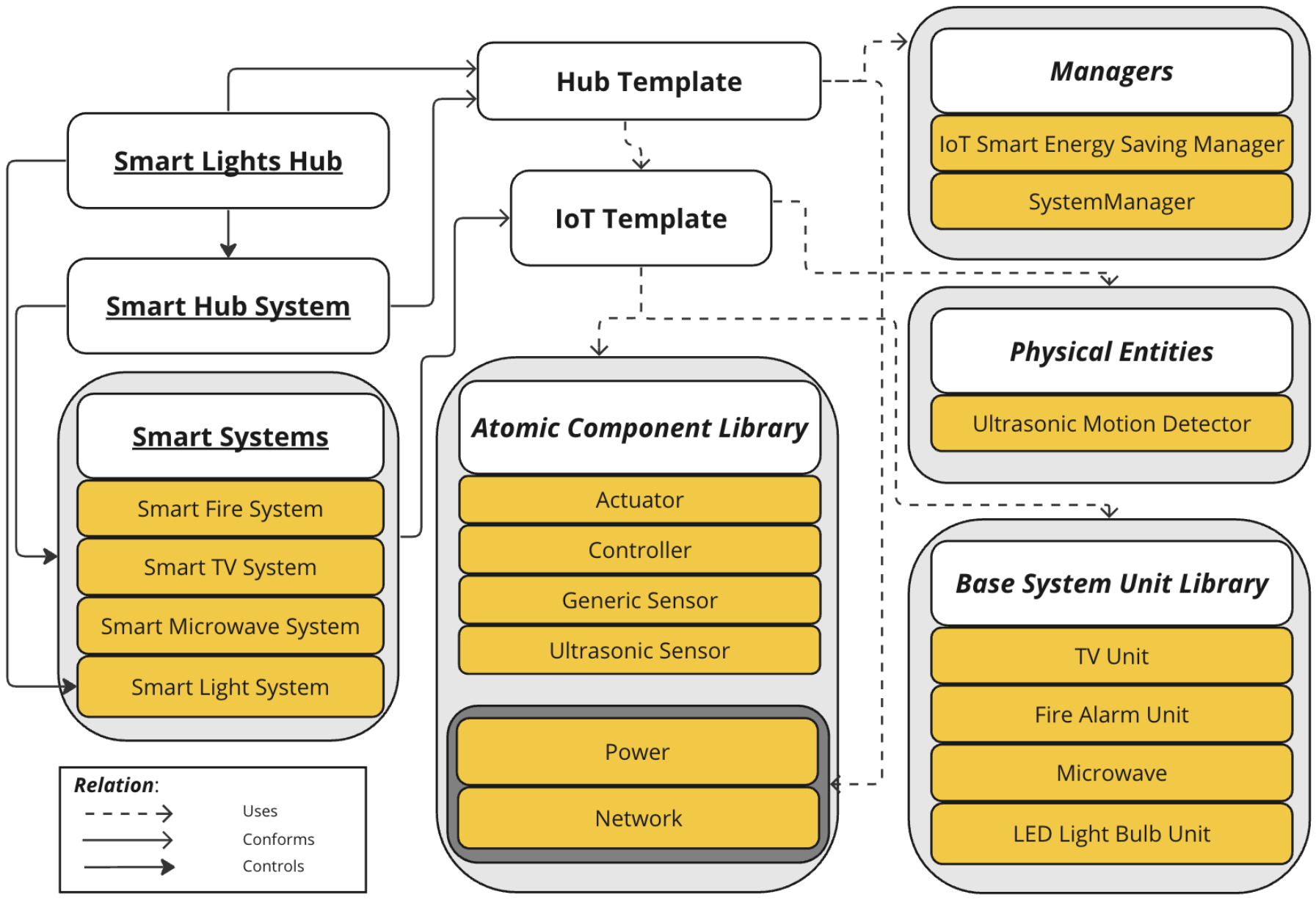

**Figure 3.** STL4IoT overview.

library is available at https://mde-tmu.github.io/iot-sc-template-library/#demo-videos.

## IoT Template

The *IoT Statechart Template* (shown in Fig. 5) contains orthogonal regions that are used to import the atomic components from the template library along with the pre-designed base system units into their designated regions. Essentially, the IoT Template allows users to design IoT systems using the pre-built templates and components. We have modelled the default template to have seven orthogonal regions, however, these can be extended depending on the system complexity. Once each system is modelled, it can then be imported into the Smart Hub Template to design a Smart Hub System, which is used to interconnect and manage other smart systems. All systems enter a default state that contains the seven orthogonal regions. These regions have composite states that contain the components with the initial state `off`. When the event `system.on` occurs, this will also raise the `on` events in all connected systems.

The `DeviceSwitchStatus`, the states `on` and `off` of the system is modelled. These transitions are triggered via the interface. The template has dedicated regions for the `Sensor`, `Controller` and `Actuator` components. We begin with the `Sensor` component monitoring by raising a timed transition event periodically (set to every 500 millisecond/ms). The `Sensor` sends a signal (i.e., a boolean activity value) to the `Controller` every 500 ms. Once the `Controller` receives a signal indicating activity, it will execute a decision-making process by coordinating with the `Base system unit` and output a signal to raise an actuating event.This enables a transition in the `Actuator`. For example, the Smart Light System will actuate the lights to turn on.

## Smart Hub Template

The Smart Hub Template (shown in Fig. 6) is composed of pre-defined orthogonal states and regions that are used to manage the coordination between multiple smart systems. Figure 6 shows the regions for each IoT system controlled by the hub. To simulate a network connection between the hub and the systems, we have also added the `Network` component in the hub.

When applying the template, more systems can be added by adding an orthogonal region per system and reusing the state machine defined in the template. The template covers some basic functionalities: 1) turning on and off individual smart systems (see first region from the left in Fig. 6), 2) turning on and off all of the smart systems at once, 3) calculating the total power consumption using the `Hub Power Manager` component, (see second region from the left in Fig. 6), and monitoring all smart system status through the hub.

The hub is modelled with the use of multi state machines. When the hub is turned on via an external event, the statechart enters multiple orthogonal regions in the `HubTemplate` state (see Fig. 6). The right-most orthogonal regions in the template are assigned to the IoT systems that are imported into the hub. These regions enter initial states and triggers an action to invoke the corresponding state machine with the `statechart.enter` event. The left-most region contains a composite state that contains substates indicating the status of the connected systems. Each system enters a default state `off` and can be turned on/off either individually or simultaneously by raising the event `HUBAllSystemsON` / `HUBAllSystemsOFF`.

## Smart Hub of Hubs (SHoH)

The concept of *Smart Hub of Hubs* (SHoH) embodies an approach within the domain of interconnected Smart Hub Systems (15; 16; 17). *SHoH* introduces a novel paradigm that goes beyond the conventional capabilities of standard smart hubs, achieving seamless integration of multiple Smart Hub Systems, each with its specific functions, within a unified and efficient ecosystem.

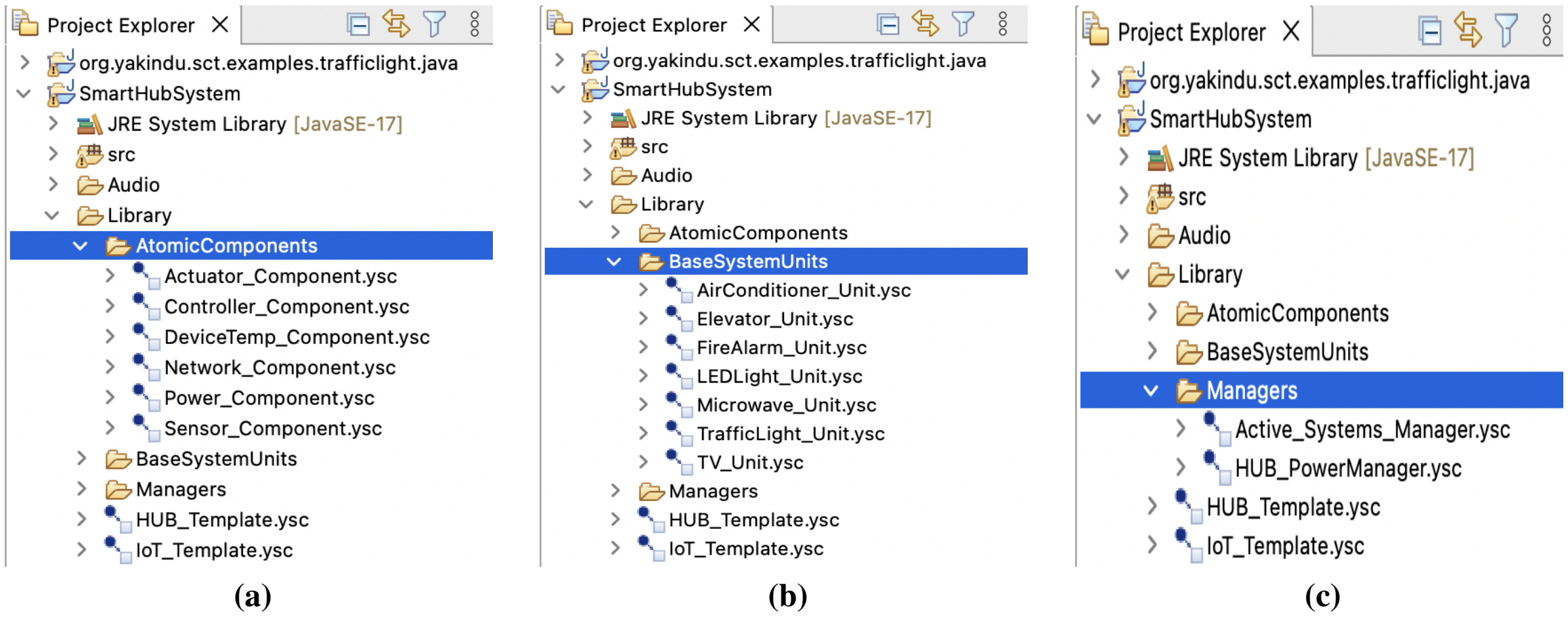

**Figure 4.** Library structure: **(a)** atomic components **(b)** base system units **(c)** managers.

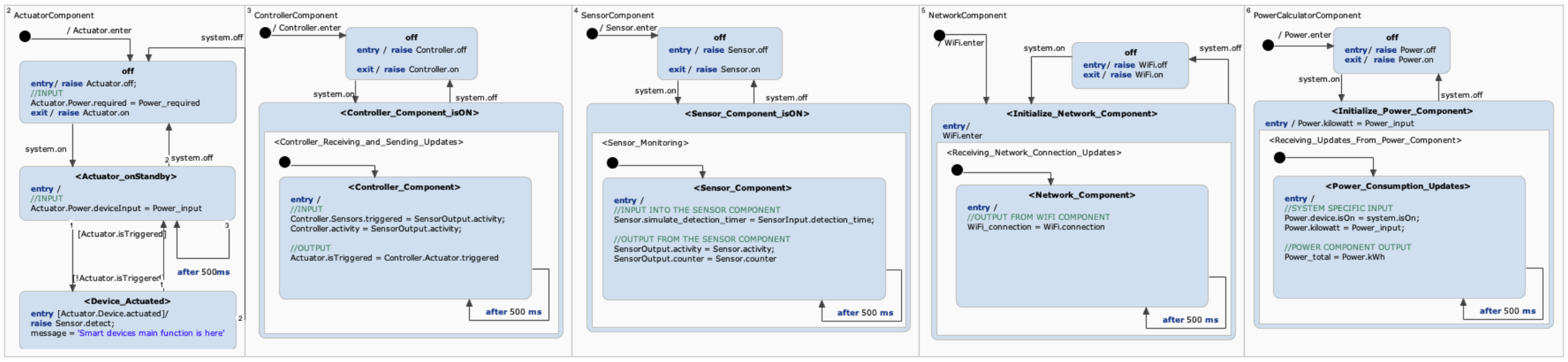

**Figure 5.** IoT statechart template (extract).

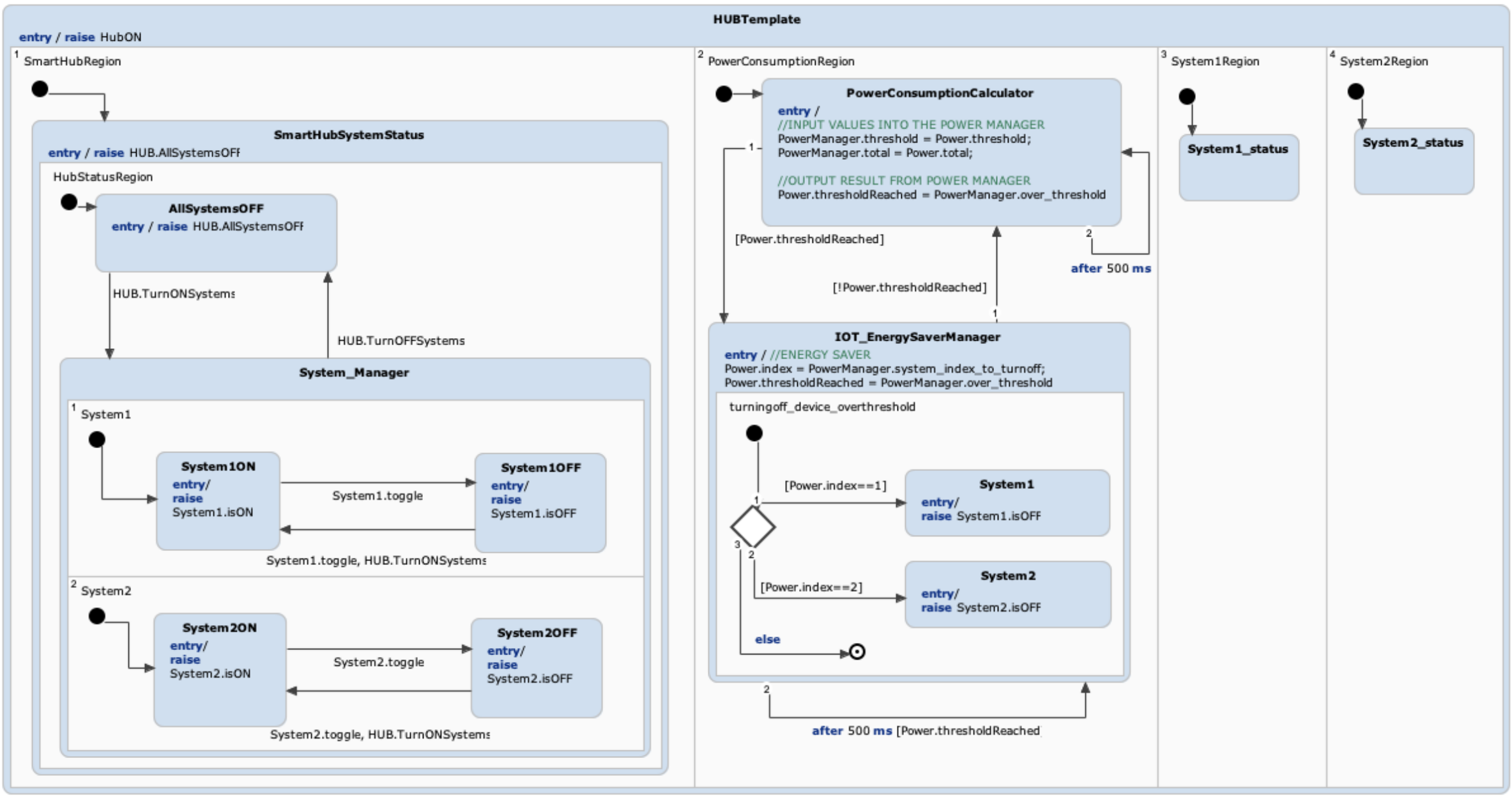

**Figure 6.** Hub statechart template.

We propose a technique to implement the hub of hubs approach using the *hub template* (see 6). We begin by identifying the hierarchical structure of the hub of hubs scenario and determining how many layers of hubs are involved and what their specific roles and functions are within the overarching system. Each layer may represent a different level of control and orchestration. Using the *hub template*, we can establish clear definitions for the relationships between the hubs at different levels and define how information and commands will flow between the hubs and how they will communicate with one another. This could involve specifying interfaces, protocols, and data exchange methods. We can do this by modelling an independent hub using *hub template* and then proceed to importing the hub into the top-level hub.

The *hub template* which is used to model hub systems also allows the integration of hubs within another hub, hence the SHoH application on Smart Lights Hub. When importing into the top-most hub layer, the hub would be added into an orthogonal region for systems and also inside `system manager` as a state. This will allow the top-most hub layer to control any number of hubs that are connected to it. For example, the *Smart Home Hub System* is able to modify and send commands to the *Smart Lights Hub* because the *Smart Home Hub System* is the higher-level hub and the *Smart Lights Hub* is connected to it.

## Atomic Components

The IoT statechart template includes a composition of components that defines an IoT smart device or system. In order to properly design an IoT system, we must model the constituent atomic components. These are designed to exhibit the behaviour of the heterogeneous software, hardware, and network components of an IoT system. We have modelled the IoT Template to be composed of the following atomic components: `device, network, sensors, actuator, controller`, and `power`. The statechart for each atomic component in the library, presented in Fig. 7, has been developed and included in our IoT library.

**`Sensor`** models the generic behaviour of sensors, which means that this sensor is not type-specific and is usable for any system. A `sensor` component is periodically "sensing" or monitoring an environment or device. As shown in Fig. 7 (e), the `sensor` component enters a default composite state `Sensor Monitoring` when the system is on. When the system turns off, the event `toggle` is raised leading to a transition to the `off` state. When in the `sensor monitoring` state, a *true* value is assigned to the *reading* variable, which indicates that the `sensor` is now reading/monitoring. The initial state, `NoActivitySensed`, indicates that there is no activity being sensed. For simulation purposes, we have set a counter value that is used to enable a *timed transition* to the `ActivitySensed` state. The `Sensor` will then send this data to the `Controller` component.

**`Controller`**, shown in Fig. 7(a), controls the behaviour of a system by receiving data from the sensors, making decisions, and raising events to trigger the `Actuators`. The `controller` will stay in the default entry state `WaitingforSensorData`, and can only transition when the conditional event `SensorsTriggered` is raised. When the `controller` receives the data from the `Sensor`, the `Controller` processes this to perform some decision making while entering the `SensorDataReceived` state. Once the `Controller` has processed the data, it will raise an actuating event, `ActuatorTrigger`, in the `Actuator` component. Once the `Controller` has communicated with the `Actuator`, it will transition to the *TriggerActuator* state until the conditional event is raised, and will then transition back to the default state *WaitingForSensorData*.

**`Actuator`**, shown in Fig. 7(b), handles the activating/actuation of devices within a system. It receives an actuation signal from the `controller` and uses this to handle the activation of the system. As shown in 7(b), the `Actuator` starts in the initial state, `StandBy`, as it waits for an *ActuatorTriggered* event from the `Controller`. Once the `Controller` sends a *true* value, then the `Actuator` will transition to the `ActuatingDevice` state.

**`Network`**, shown in Fig. 7(d), handles the gateway connection between the system and the network. Each system may have a different network type such as *cellular, Wi-Fi, LPWAN, bluetooth* and *Zigbee*. Currently, only the statechart for Wi-Fi is included in the library. The `Network` component starts in an initial state `NetworkComponentWorking`, and can transition to the `off` state by raising the `off` event. Otherwise, it transitions and enters a composite state `checkingForNetworkConnection` and the default substate `connectingToServer`, which will indicate either a *successful* or *failed* connection using a conditional event. Additionally, for simulation, we have modelled the component to have a periodic network timeout.

**`Power`** is an atomic component that handles the per-hour power consumption (in KWh) of the system, as shown in Fig. 7(c). It calculates the total consumption of the system and this data is sent to the `Hub Power Manager` through the `Smart Hub System`. It contains a composite state the describes the consumption status of the system. The statechart has a default entry state *NoPowerConsumed* as each system always starts in the *off* state. Once the system is turned on, this will also reflect in the `power` component and raises the conditional event. The substate will then transition to the `ConsumingPower` state as it calculates the per hour consumption using the kilowattsPerHour(kWh) units.

## Base System Units

The **Base System Units** library, shown in Fig. 4, is a repository of pre-designed units that can function autonomously without inherent smart capabilities. These units can be imported from the library and are to be treated as self-contained systems that can function on their own but does not contain any specific "smart" functionalities. The infusion of smart features is achieved by integrating the atomic components provided by the IoT library. The objective is to gradually expand this library by accumulating a repository of reusable units.

## Physical Entity

**`Physical entity`** components play a crucial role in modelling smart systems, as they represent the tangible and often sensor-equipped elements within the IoT ecosystem. These components, designed using the existing atomic components available in the library, offer a template (see Fig.

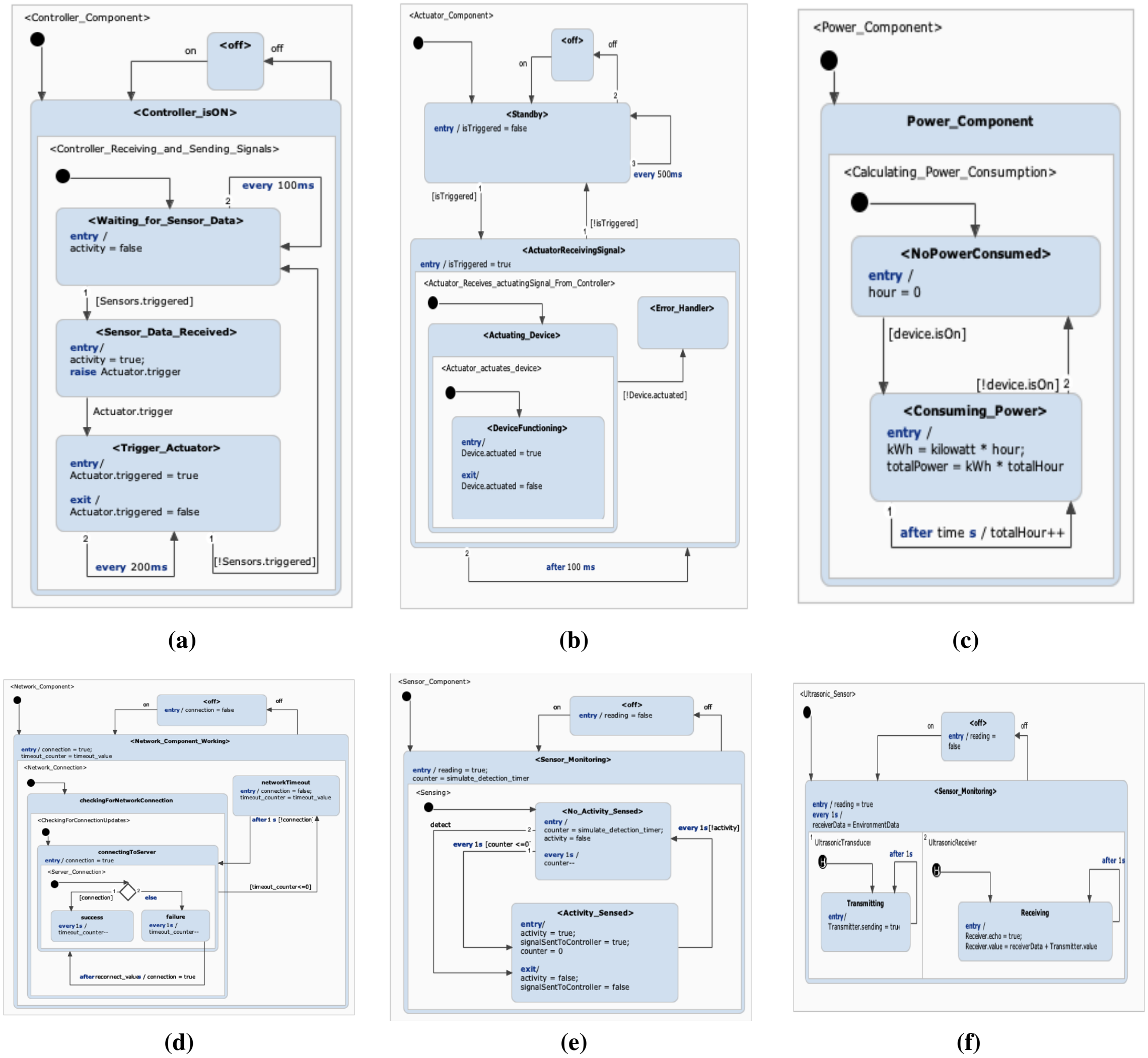

**Figure 7.** Atomic components statechart templates: **(a)** controller **(b)** actuator **(c)** power **(d)** network **(e)** generic sensor **(f)** ultrasonic sensor.

8) for modelling the behaviour of physical devices, such as sensors, actuators, and other system.

Within the `physical entity` statechart, we have regions specified for the atomic components such as `sensors`, `controller`, `actuator` and `power` components. The template allows designers to model physical entities with existing components in our library and to create a new component that can be utilized in smart systems. For example, we have modelled a *Ultrasonic Motion Detector* physical entity that uses an ultrasonic sensor from the STL4IoT library. This entity also uses controller and actuator components which play roles in the behaviour of the motion detector as the controller receives monitoring data from the sensors. The sensors continue to read data as the controller simultaneously carries out the decision-making process and trigger the actuator.

## *Manager Components*

The **Manager Components** are pre-designed components that contain special manager features that can *only* be imported into the Smart Hub Template. An example of a managing component is the `Power Hub Manager`. The **`Power`** component designed within the IoT systems calculates the system's power consumption as shown in 7(c) and sends data to the hub. Whenever a device changes its status (e.g., from `on` to `off`), the total power consumption is updated immediately. Once the power threshold is reached, the `Hub Power Manager` transitions to the state `CheckingDevicePowerContribution` to check which device contributes the most to the total power consumption. Once the device consuming maximum power is identified, the hub will turn it off. This can be observed in Fig. 9.

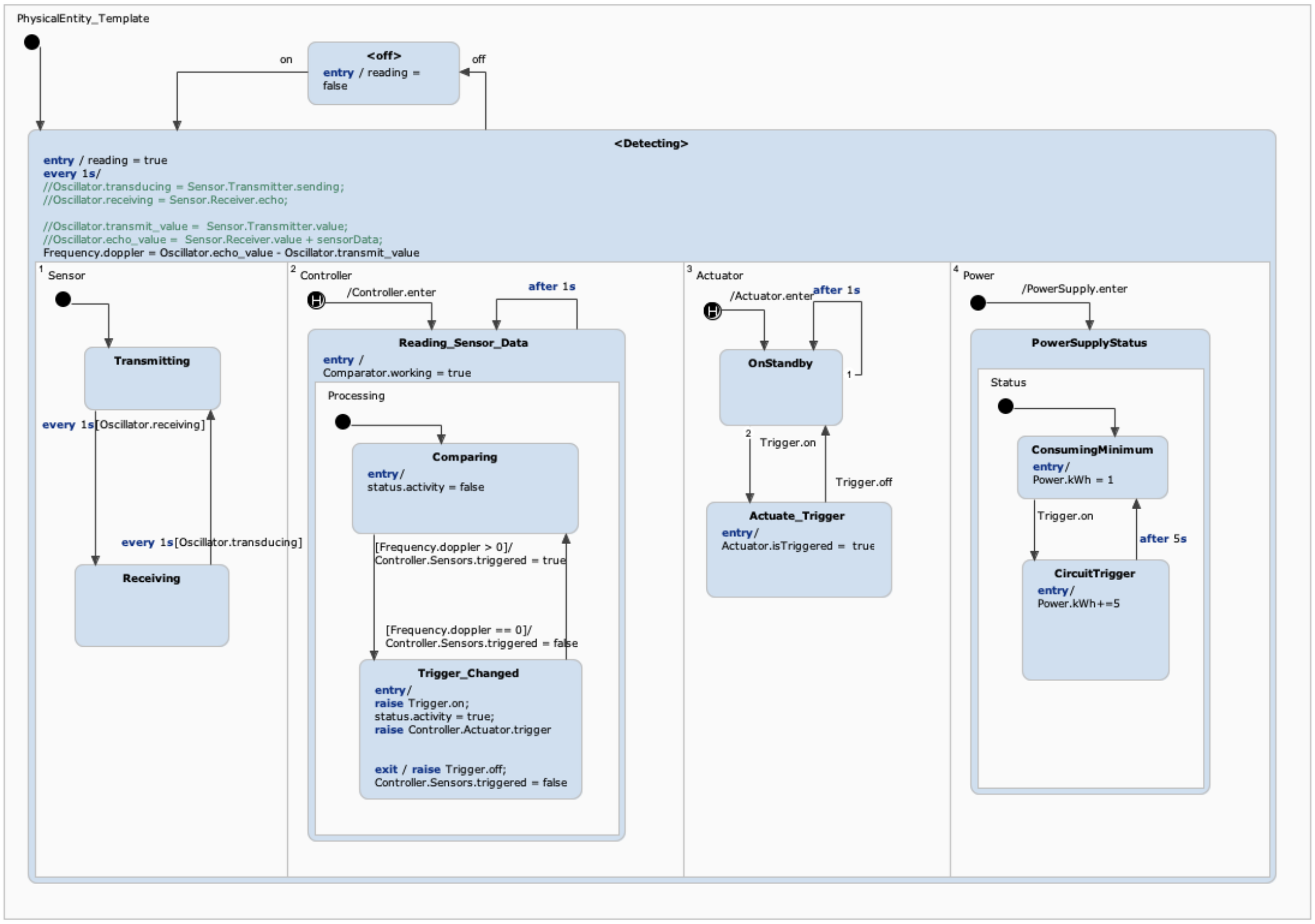

**Figure 8.** Physical entity template.

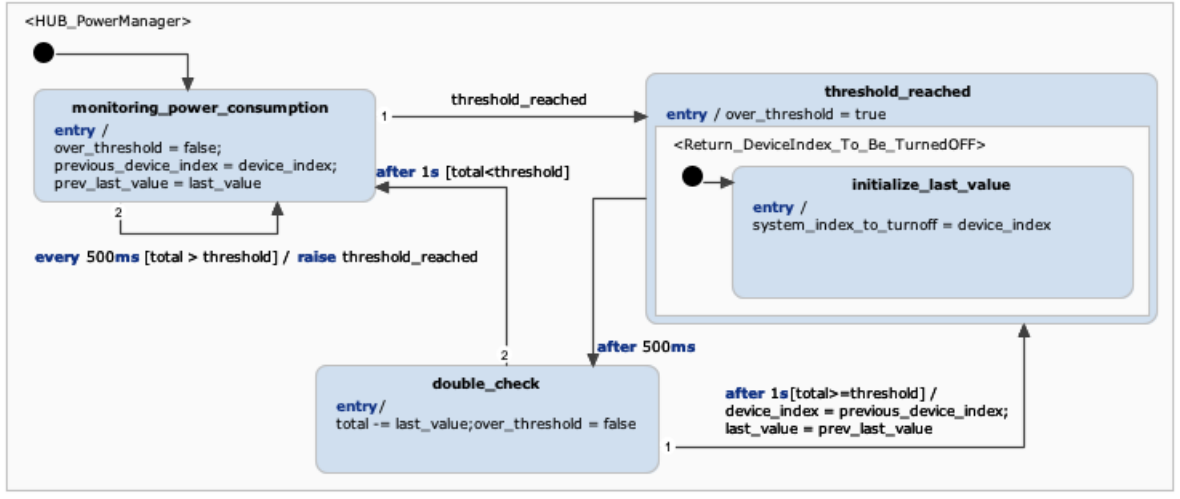

**Figure 9.** Manager component: hub power manager.

## Modelling the Smart Home System Controller

In this section, the proposed library of templates is used to model the smart home system, introduced in Sec. . Using the *IoT Template*, we have designed multiple smart systems and connected all the systems to a smart home hub by designing a main controlling hub using the *Hub Template*.

### *Using the Templates*

To use these templates, the user must be able to understand the structure and design of the templates. In the IoT template, we have designed it to contain all of the necessary atomic components (ie. `sensor, controller, etc.`). Every region and composite state are pre-configured with the variables, events, and transitions, hence, no changes are required in the atomic components. The following are guidelines for modellers using our library of templates.

1. Modeller must refer to the *IoT Template* when designing a smart system. If the modeller desires to model a hub to link smart systems together, then the *Hub Template* must be used.
2. Once a template is chosen, the modeller must determine which states and regions are editable.
3. For the *IoT Template*, the modeller can only edit the model by adding more `atomic components` and the `base system units`. The user may refer to the `atomic components` library and import any additional components.
4. The application-specific functionality has to be added as a *base system unit*. The modeller can refer to the *base system unit* library for this, however, if the desired unit is not available, then the user will have to design their own. Adding a unit means importing the *base system* statechart via the configuration panel, and then assigning an appropriate variable to the statechart. This will allow for an easier use of the *base system*'s built-in variables and events. Once the *base system* and the *components* are finalized, the statechart can now be used as a smart system.
5. For the *Hub Template*, the modeller only needs to edit the number of smart systems connected. The hub can be connected to numerous smart systems as long as it is imported and extended properly. This can be done by importing the smart system statecharts via the configuration panel and assigning these to specific system variables. Doing this will allow accessibility to the smart system variables and events from the Smart Hub level.
6. Using the *Hub Template*, the designer can also model a SHoH system by defining each hub layer and integrating a hub into a higher-layer hub.

A systematic approach to use the hub template for the designing a ”Smart Hub of Hubs” (SHoH) is described below.

1. Identify the multiple Smart Hub Systems that you want to integrate within your environment, each with its specific functions. The hubs to be composed are not required to be designed with the use of the `Hub Template`
2. Create a centralized hub using the `Hub Template`, referred to as the “Smart Hub of Hubs”, which will serve as the orchestrator for these interconnected hub systems. You can import each hub system just like how you would import individual smart systems.
3. Utilize the hub template to design the architecture of your Smart Hub of Hubs. This template should define the relationships and communication protocols between the various hub systems, ensuring seamless integration and efficient operation.
4. Once the hub design is established, you can proceed to implement it by connecting the individual instances of smart lights or other devices to the respective Smart Hub Systems and, in turn, linking these systems to the Smart Hub of Hubs.

In the remainder of this section, we present the statechart models designed for the smart home application with the use of our template library.

## Smart Home Design

The *Smart Home* statechart, designed using the *Hub Template*, aims to connect and link all smart systems located within a smart home system as shown in the orthogonal regions in Fig. 10. Once the hub system enters the *on* state, all of the orthogonal regions will also enter all of the corresponding initial states. For example, in the left-most region of Fig. 10, the initial state indicates that all systems, excluding the smart fire system, will enter the `off` state. The hub system has a *PowerManager* component, which basically monitors the total power consumption of the system (see second left region of Fig. 10. The power manager continuously monitors the total power and only exits from the `PowerConsumptionCalculator` state when the event `PowerThresholdReached` is raised in the `HubPowerManager`. Once this event occurs, the state transitions to *IoTEnergySaverManager*. In this state, it receives the system index output by the `HubPowerManager` and turns off the system that matches the index value.

## Smart Fire Alarm Design

The Smart Fire Alarm System is composed of the main atomic components, namely `sensor`, `actuator`, and `controller` (see Fig. 11). We have specifically added three sensors for carbon, smoke and heat. We use the `IoT Template` as the main skeleton for system design.

The basic functionalities of a smart fire system are pre-designed within the base *Smart Fire* unit and is imported into the template (see Fig. 11). Once imported, the predefined atomic components will work accordingly. For simulation purposes, random detection times are assigned to the sensors (via the UI). Once the detection occurs, this raises an event that enables a transition to the *ActivitySensed* state within the `Sensor` component. The `Controller` will then process the input from the sensor in the *base system unit* statechart and monitor the sensor levels by raising the *SensorTriggerSignalReceived* event and transitioning to the *SensorTriggered* state. The system will also enter a `warning state` by raising the *Warning* event until the threshold value for the `sensor` is reached. Once the threshold is reached, the `controller` will send an actuation signal to the `actuator` to sound the alarm by raising the *Danger* event, which sends the *base system* to `Danger` state. When the `actuator` receives the actuation event, the fire alarm is triggered. Furthermore, the system will automatically send an emergency signal event to the `Smart Hub System`, which would send the hub to the `Emergency` state. Consequently, the `Smart Hub` will handle this emergency event and restrict all other systems within the environment from being used in the `EmergencyState`. This can only be turned off once the fire alarm is turned off through the Smart Fire System.

## Smart TV Design

The Smart TV System is composed of the main atomic components such as the `sensor`, `actuator`, and the `controller`. The `IoT Template` was used to design the root statechart of the system, while the specific TV functionalities are defined and described within the `Base TV unit` as shown in 13. Once the `base TV unit` is imported, the other predefined components will work accordingly.

The base TV unit statechart works independently without the *smart* components. It models the core TV functionalities such as changing channels and managing the input sources. Additional *smart* components such as the *WiFi Gateway*, *Power Calculator*, and *Temperature Monitor* are integrated within the template. Unlike the *Smart Fire System*, we use the Smart TV’s *sensors* specifically for detecting inactivity. The intention is to save energy when the system is not being used. The *sensors* will continue to monitor for activity in the near surroundings. When there is a lack of activity, the `sensor` component will transition to the active state. The sensor data is then received by the `controller` and processed within the *base system unit* (TV unit in this case). The base system controller will process this and raise an actuation event in the `actuator`. In the `actuator` component, the actuatorTriggered event is then raised, which results in a transition to the `DeviceActuated` state. Once the smart system is in the `DeviceActuated` state, this indicates that the TV has been put to the *sleep mode*, since the system’s main goal is to save energy when not being used. Similar to the Smart Fire System, this system will be restricted from being used by the smart hub when in an emergency mode.

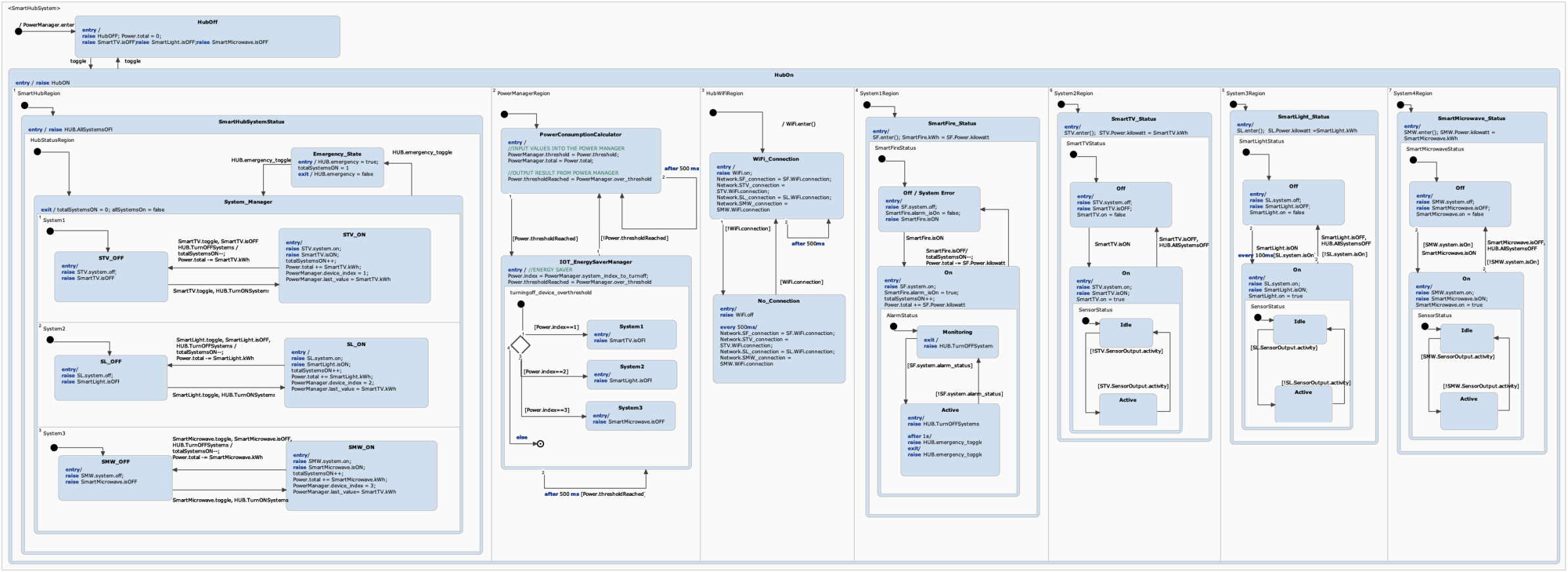

**Figure 10.** Smart home system: statechart model.

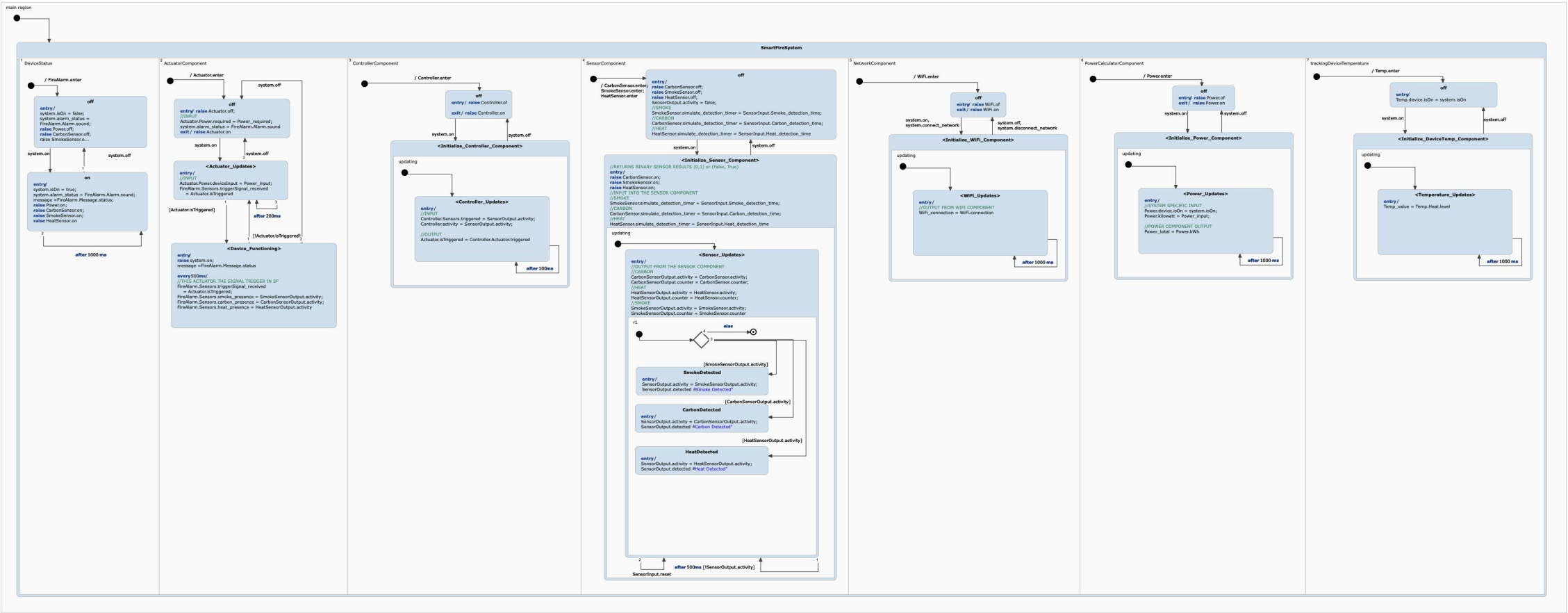

**Figure 11.** Smart fire system: statechart model.

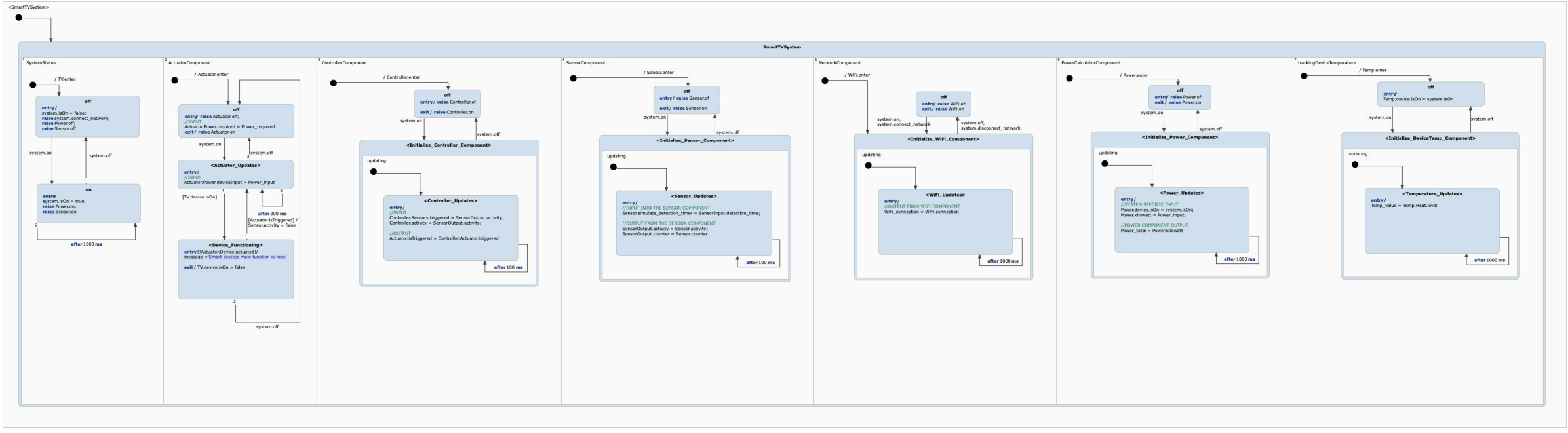

**Figure 12.** Smart TV system: statechart model.

## *Smart Microwave Design*

The Smart Microwave System is a smart system that has the traditional microwave functionalities and some additional smart features. Similar to the previously discussed systems, it is composed of the atomic components, `sensor`, `actuator`, and `controller`. We used the `IoT Template` to design the skeleton of the system (see Fig. 14). The `Base Microwave unit` is imported into the `IoT template` along with the `atomic components`. The `base microwave unit` works autonomously without any smart features as shown in 15. It includes the basic microwave functionalities such as `door status`, `start/end timer`, and `add time`. Integrating the `base microwave unit` in the `IoT Template` with the `atomic components` extends the base design to that of a smart microwave with the addition of *smart* features, such as the food and heat sensors in the microwave. For this instance, the `sensors` are used to sense presence of food on the microwave plate using its' weight. For the food detection, the `sensor` allow the microwave to function and start the heating up process. Lack of weight on the microwave plate means that the `sensor` component is in the `NoActivitySensed` state, as activity for this instance indicates "food presence". This will result in the system restricting the user from starting the timer by returning an invalid weight value. Until the `weight sensor` returns a valid value,the system will remain in the `StandBy` state. Once the `sensor` returns a valid weight value, the `controller` will process this and raise an event in the `actuator`. The system also has a `Door Status` that monitors the microwave door. If the door status is changed then the system will send the signal to the `controller` to raise a false-boolean guard event in the `actuator` component and pause the heating process. Essentially, the system will recognize this and stop the timer to inform the user about the issue by raising the `pause` event in the `base system unit`.

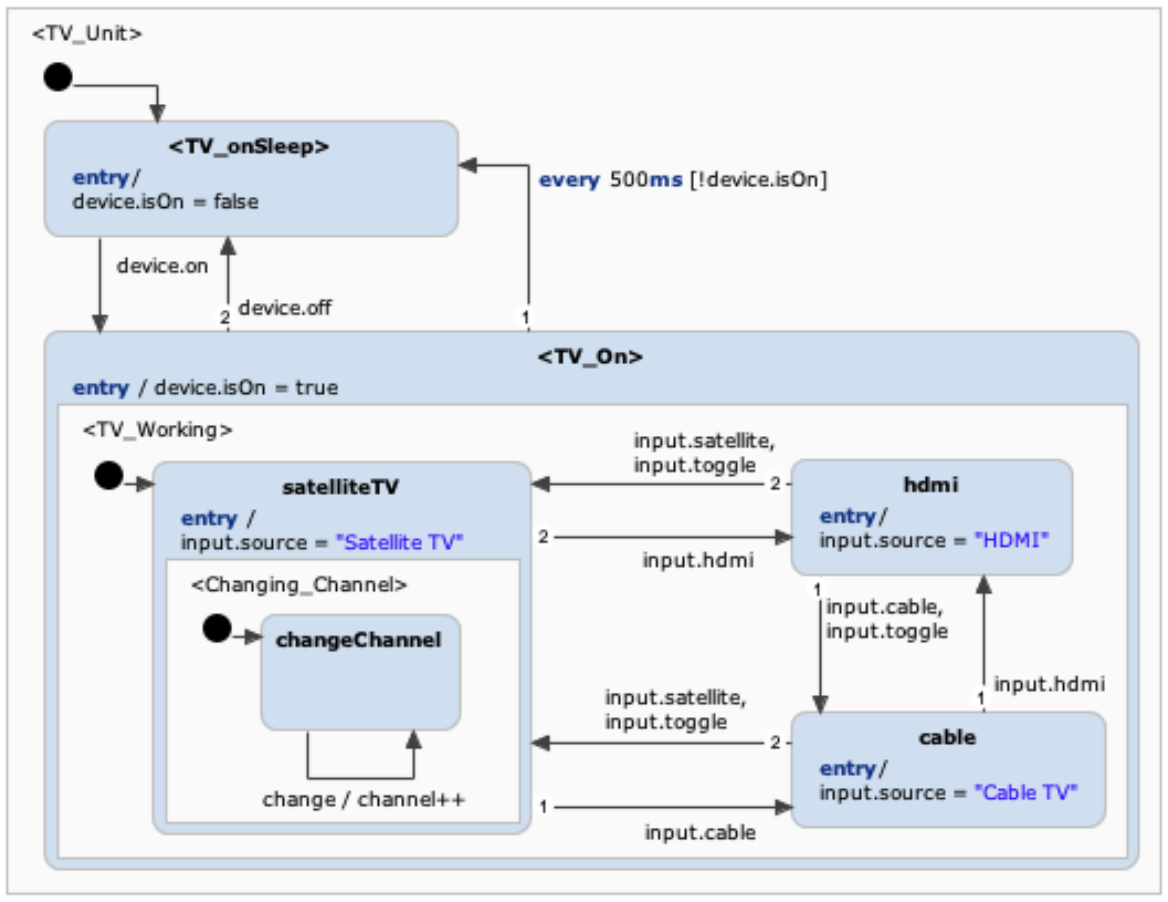

**Figure 13.** Base TV unit.

### *Smart Lights Design*

The Smart Lights System is also implemented with the atomic IoT components, `ultrasonic sensor`, `actuator`, and `controller` from the library. The basic light functionalities are defined within the `Base LED Lights unit` and is imported into the template. The base unit has the main functionalities such as `on/off`, `dim/brighten lights`. The Smart Light System's `sensor` monitors the environment for activity. Once activity is sensed, the `Sensor`'s activity data is received by the `Controller` component. The `Controller` in the base system unit recognizes and processes this data, and then raises an actuating event to turn the lights on. Additionally, once the `Sensor` outputs a false value, the `Controller` will raise another actuating event and transition back to the `standby` state.

### *Smart Hub of Hubs (SHoH) Design*

The successful application of the hub template at the level of the Smart Home Hub of Hubs, as demonstrated in Fig. 18, validates the use of the STL4IoT approach within the domain of smart hubs. While further testing and refinements may be necessary to ensure versatility and compatibility with specific use cases, this approach provides a powerful framework for managing interconnected Smart Hub Systems and optimizing their performance within a unified ecosystem.

## Smart Home Simulator

### *Simulation Dashboard*

For our training simulator, we have developed a dashboard with Java that serves as the user interface (UI) for the smart home hub. Using the code generation feature of *Itemis Create*, we generated Java source code from the statecharts, which is then integrated with the user interface. The source code is regenerated with every change made to the statechart.

In order to effectively display and plot data, we have made use of Java and Java libraries, JFrame[*] and JFreeChart[†], enabling the visualization of data extracted from statechart's source code. Integration of plots in the dashboard enhances the understanding of system behaviour and facilitates informed decision-making in IoT system design. A video demonstrating the smart home system developed with our templates is available at `https://mde-tmu.github.io/iot-sc-template-library/#demo-videos`.

The dashboard is composed of 9 panels as shown in Fig. 19.

- Panel 1 in the top-left displays the system status of hub, which includes its' network status, notification system, master switch, main hub switch and the power manager.
- Panel 2 or middle-top panel indicates the status of all of the smart systems connected to the hub, such as its' usage, power, and network connectivity status. The status can be viewed by clicking on the tab button.
- Panel 3 or top-right panel shows the percentage of the power consumed by each system in relation to the current total power consumption. The pie chart includes systems that are turned on and are contributing to the power consumption.
- Panel 4 or middle-left panel displays the status of the Smart Fire System. It shows a bar graph depicting the carbon, smoke, and heat levels based on sensor readings.
- Panel 5 or the middle panel contains the control system for the Smart TV System.
- Panel 6 or the middle-right panel shows the total power consumption of the home environment.
- Panel 7 in the bottom-left contains the control system for the Smart Lights System, which is essentially a smart hub of lights.
- Panel 8 or the bottom-middle panel contains the control system for the Smart Microwave.
- Panel 9 in the bottom-right is the report logging panel which logs all the reports and saves it into a data text file.

[*] https://www.javatpoint.com/java-jframe
[†] https://www.jfree.org/jfreechart/

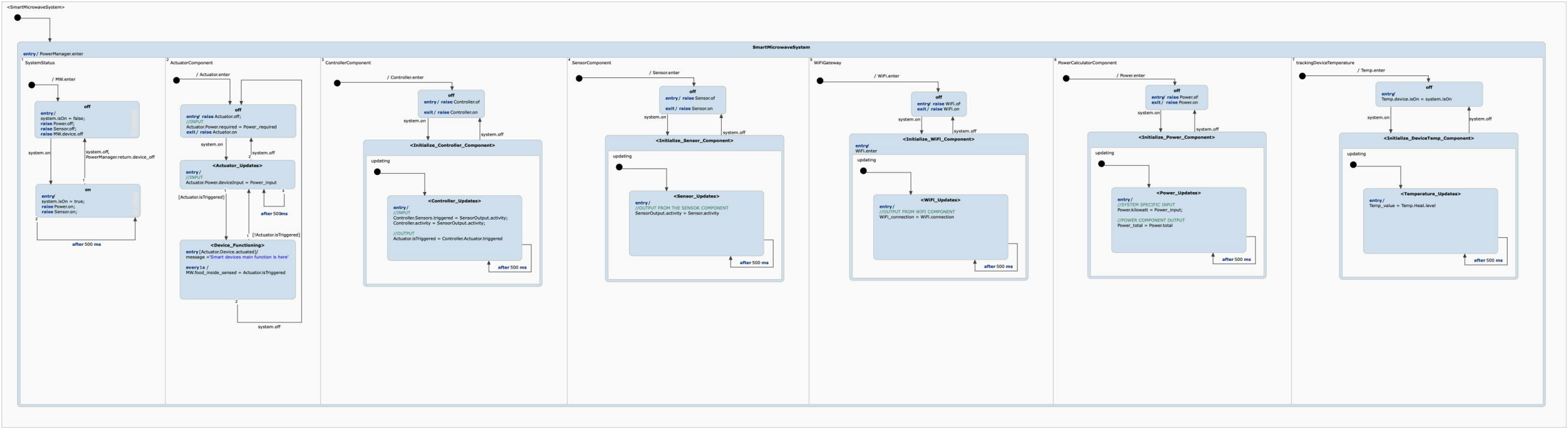

**Figure 14.** *Smart microwave system: statechart model.*

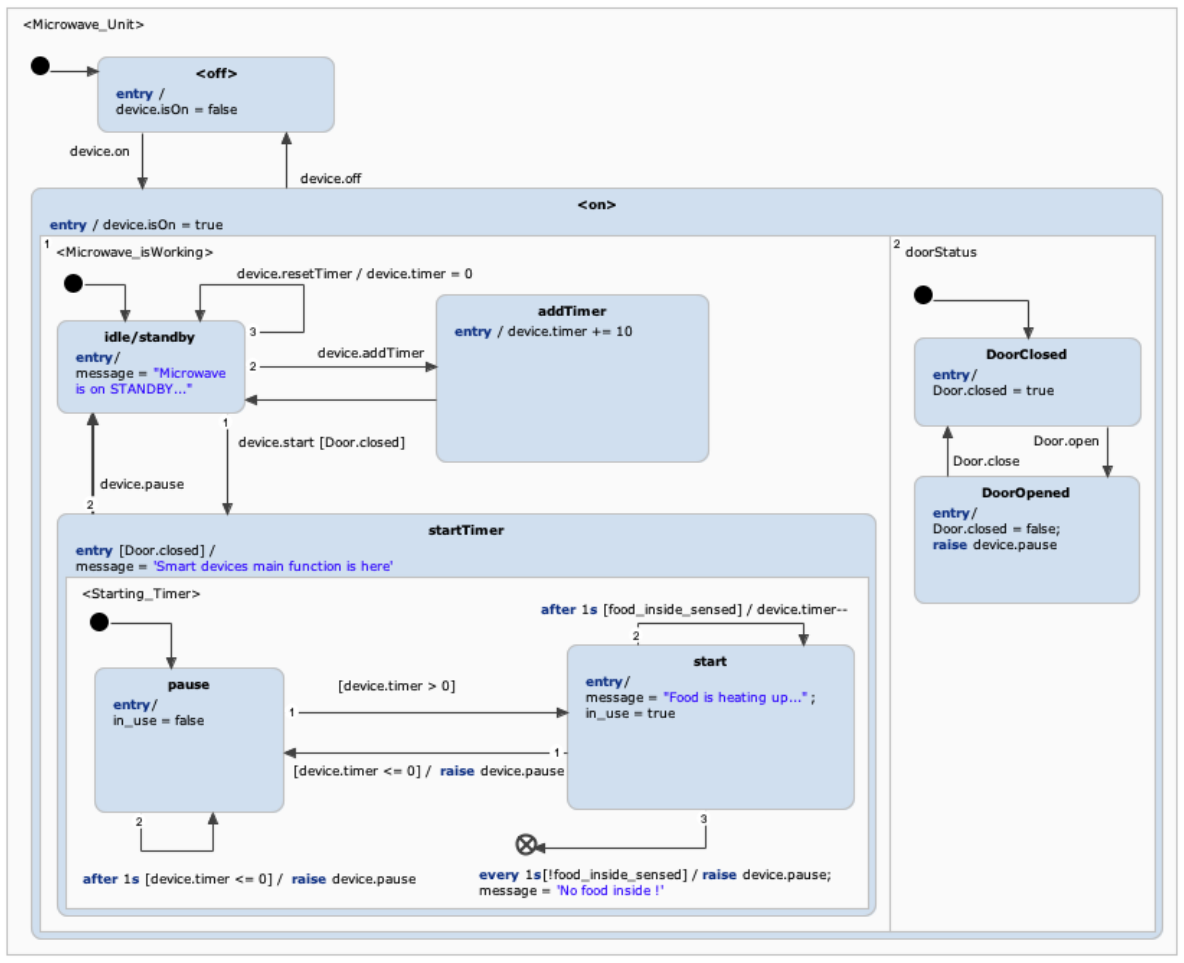

**Figure 15.** *Base microwave unit.*

Moreover, we have implemented separate testing panels as shown in 20. The testing panel is implemented to test the functionality of the simulator through the UI. Each button is used to simulate the process of triggering each specific sensor. For example, in the `Smart Fire Testing Panel`, there are 4 buttons: one each for the sensors integrated in the *Smart Fire System* and one button for triggering the fire alarm system. Instead of waiting for the system to progress into a danger state, the testing panel provides the ability to test the system functionality once a sensor senses a danger presence.

### *Simulation Environment*

We have also explicitly modelled the simulation environment which is responsible for generating sensor readings and configuring the data variables for each smart home system. The randomly generated sensor values are logged and saved into a data file and is then read by the sensors of each smart system.

### *Simulation Results*

We have implemented data plotting to present the simulation results for power consumption (see Fig. 19). The middle-left panel displays the change in carbon, smoke, and heat levels in the home based on the sensor readings in the Smart Fire System. The top-right panel shows the power contribution of each smart system with a pie chart, while the middle-right panel shows the total power consumption of the smart home.

## Related Work

In this section, we discuss existing research work and projects on modelling and simulation-based design of IoT and CPS systems.

### *Template-Based and Composition Frameworks*

Molnar et al. (18; 19) introduce the Gamma framework for development, validation, and verification of reactive systems. Gamma uses statechart models for designing atomic components, and system assembly is facilitated by a domain-specific modelling language, *Gamma Composition Definition*, built on top of itemis CREATE. Gamma provides statechart composition support with the use of a textual scripting language but does not offer any templates for statechart modelling. Csuvarszki et al. (20) builds on the Gamma framework (18; 19) by extending its' composition semantics. This work extends the use of Gamma to the CPS domain. STL4IoT provides a template library as well as composition support based on statecharts and is tailored for IoT system design.

It is worth noting that there is a notable lack of work on statechart-based frameworks and templates, especially within the IoT/CPS domain. Apart from statechart-based approaches, Dinkelbach et al. (21) introduces a methodology for the automated generation of programming language-specific code from CIM/CGMES specifications by leveraging a template language to ensure complete compliance with CIM/CGMES specifications within software projects. They detail the process of code generation and the integration of codebases. They provide practical examples, including the development of a CIM/CGMES web editor in JavaScript and two CIM/CGMES libraries in C++ and Python. The proposed approach is tailored for Smart Grids.

Chatterjee et al. (22) proposes a modelling tool, xMAS, which deals with the micro-architectural level of system design, particularly in the context of communication fabrics. It identifies a set of micro-architectural primitives to describe complete systems through composition alone, aiming to simplify the modelling process and reduce common errors. While STL4IoT has a similar goal, it is domain-specific

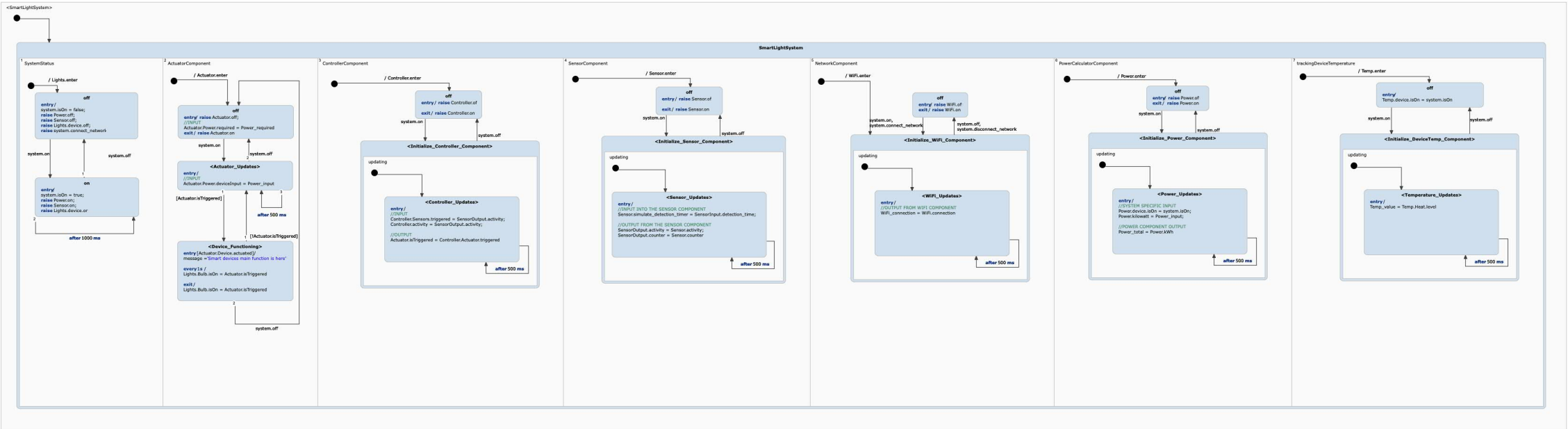

**Figure 16.** Smart light system: statechart model.

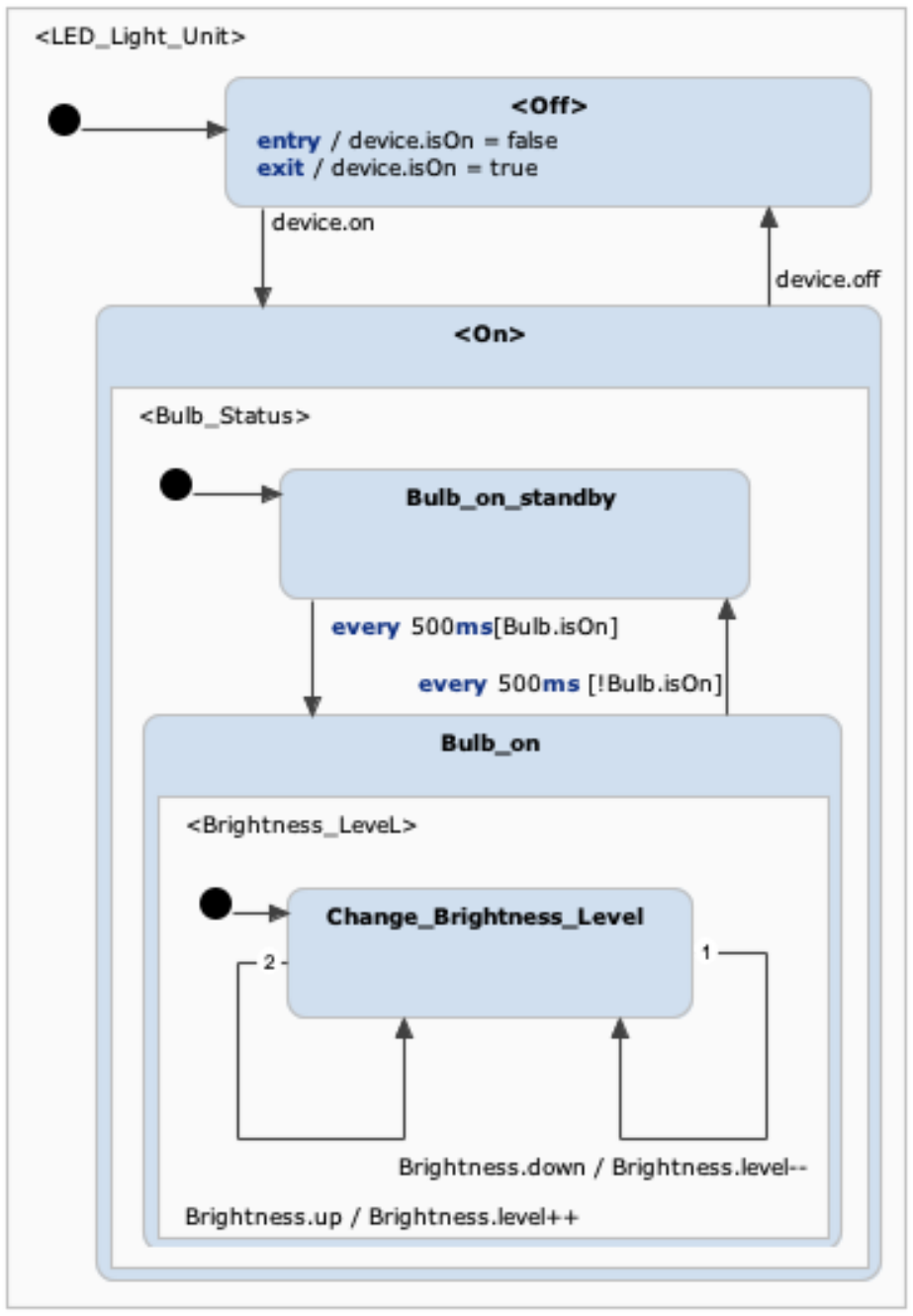

**Figure 17.** Base LED bulb unit.

and the library was developed by integrating IoT and CPS domain concepts and components.

## Modelling and Simulation Approaches for IoT and CPS Design

In this subsection, we discuss existing research work pertaining to modelling and simulation for IoT and CPS application domains, albeit without the use of templates. We considered approaches that use statecharts as well as other modelling languages.

D'Angelo et al. (3) emphasizes the significance of simulating IoT scenarios for evaluating intelligent service deployment strategies across diverse territories. Tackling the complexity of these scenarios, this work aims to create scalable simulation environments capable of real-time execution in highly populated IoT ecosystems. Combining these techniques with agent-based, adaptive Parallel and Distributed Simulation (PADS) approaches enables detailed simulations as needed. The approach is demonstrated with the use of a vehicular transportation system.

Sztipanovits et al. (23) proposes OpenMETA for CPS and aims to reduce design cycles through the "correct-by-construction" principle. While OpenMETA aims for horizontal integration layers to enhance flexibility and adaptability in CPS design flows, STL4IoT provides a structured template and components for modelling and simulating diverse IoT applications.

IBM'S RSARTE (24) serves as a comprehensive modelling and development environment specifically aimed at creating stateful, event-driven real-time applications, particularly in safety-critical contexts like complex cyber-physical systems. There is, however, no provision of templates for design or composition of CPS.

Boutot et al. (25) introduces IoTMoF, a model-driven framework for rapid prototyping of IoT systems. IoTMoF provides support for requirements development, platform-specific modelling, and code generation for IoT systems. It employs a domain-specific use case modelling language, UCM4IoT, for requirements modelling, along with a domain modelling language aligned with the IoT Architectural Reference Model (ARM) (12). IoTMoF and STL4IoT share a common goal of facilitating the design and development of IoT systems through model-driven approaches. Both frameworks use statecharts for behavioural modelling. However, IoTMoF does not provide or use any templates or patterns for statechart modelling. The work is applied on a simple smart light application and does not address composition challenges of complex IoT systems.

Xiao et al. (26) proposes a coherent architecture for IoT development. Their objective is to enable interoperable, low-cost, and user-customizable IoT rapid prototyping. Within this architecture, each IoT component is abstracted into an independent web service described by transferable states. The research establishes a Finite-State-Machine (FSM) model-driven architecture, exemplified by the Hyper Sensor Markup Language (HSML). Almeida et al. (27) proposes an approach for developing multi-modal multi-device applications using SCXML state machines, focusing on interaction aspects or behaviour within multi-device systems. Sinha et al. (28) introduces parametric statecharts, an extension of statecharts that can be dynamically customized to accommodate varying configurations of smart homes.

STL4IoT goes beyond the modelling support provided in these approaches by providing reusable templates and components that enable IoT designers to model without having to start from scratch, thereby promoting easier access to fast and easy modelling of smart systems using the templates and component library we have provided.

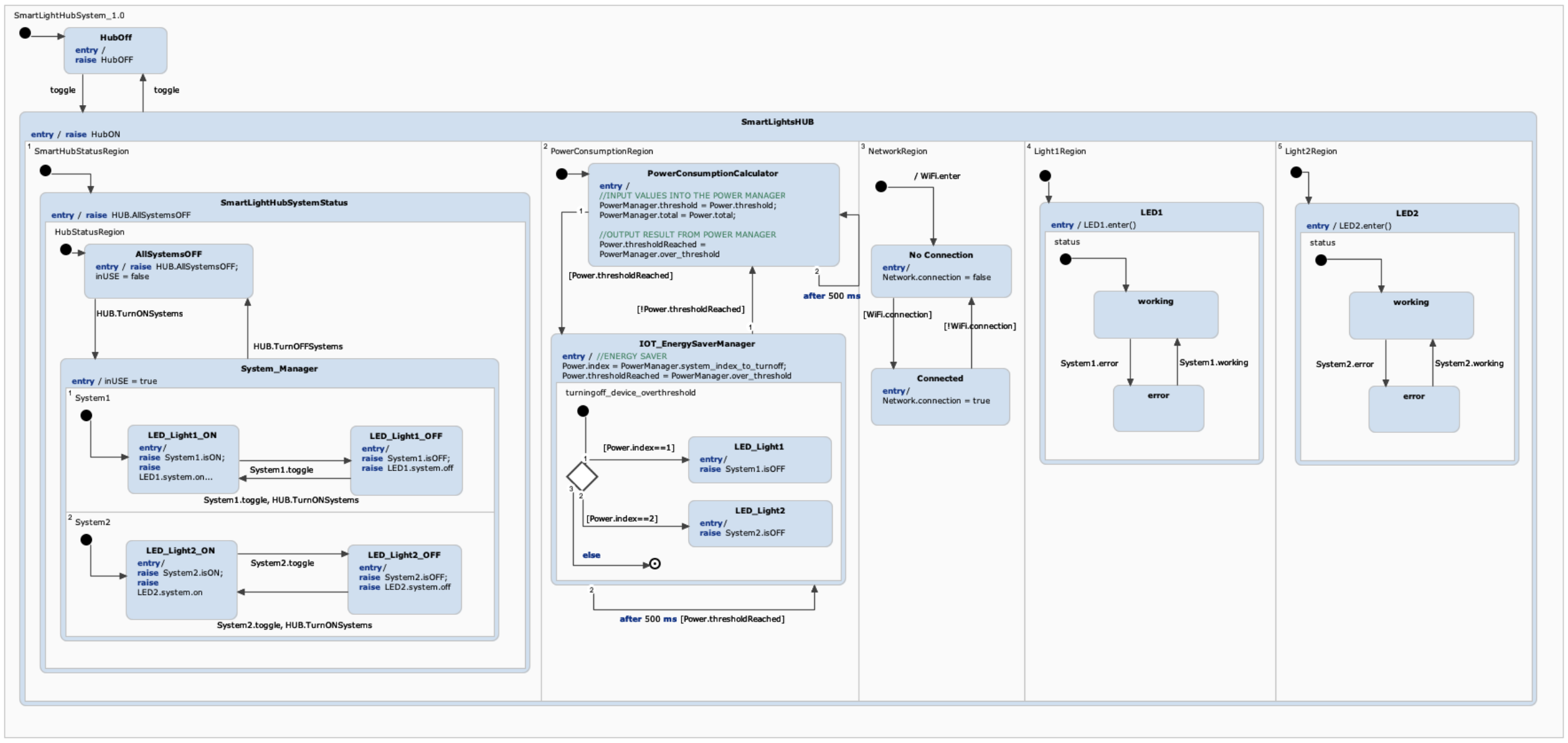

**Figure 18.** SHoH example: Smart lights hub statechart model.

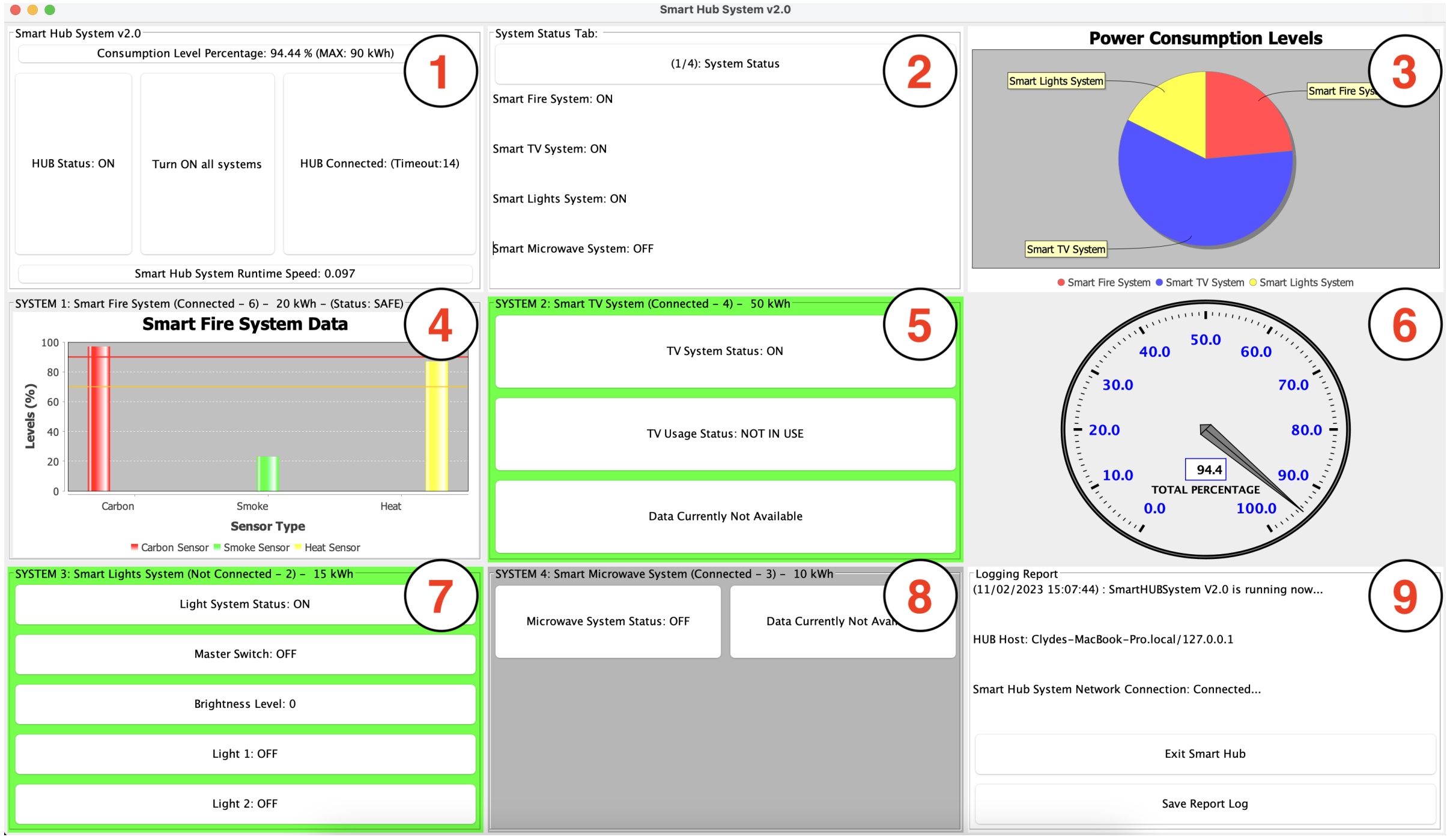

**Figure 19.** Smart home system v2.0: simulation dashboard with plots.

## *Smart Home Design and Development*

There exists a wide array of work on smart home automation focusing on IoT components, hardware platform, and implementation aspects or on the use of artificial intelligence (Williams (11), Gunge (2)). These work have helped us understand the architecture of IoT systems and derive the core components to integrate in our templates. In the context of modelling and simulation, there is some existing work specifically for smart home design and automation.

Costa et al. (29) focuses on the challenges of representing heterogeneous entities and verifying Quality of Service (QoS) properties in IoT applications. The proposed approach involves SysML4IoT and SysML2NuSMV, utilizing a SysML profile based on the IoT-A Reference Model and a model-to-text translator for model checking.

Harrand et al. (30) presents ThingML which comprises a modelling language, tools, and a methodology for developing IoT applications. The ThingML language integrates software-modelling constructs aligned with UML, including statecharts and components, along with an imperative platform-independent action language and constructs tailored for IoT applications. The tools include editors, transformations, and a multi-platform code generation framework supporting various programming languages.

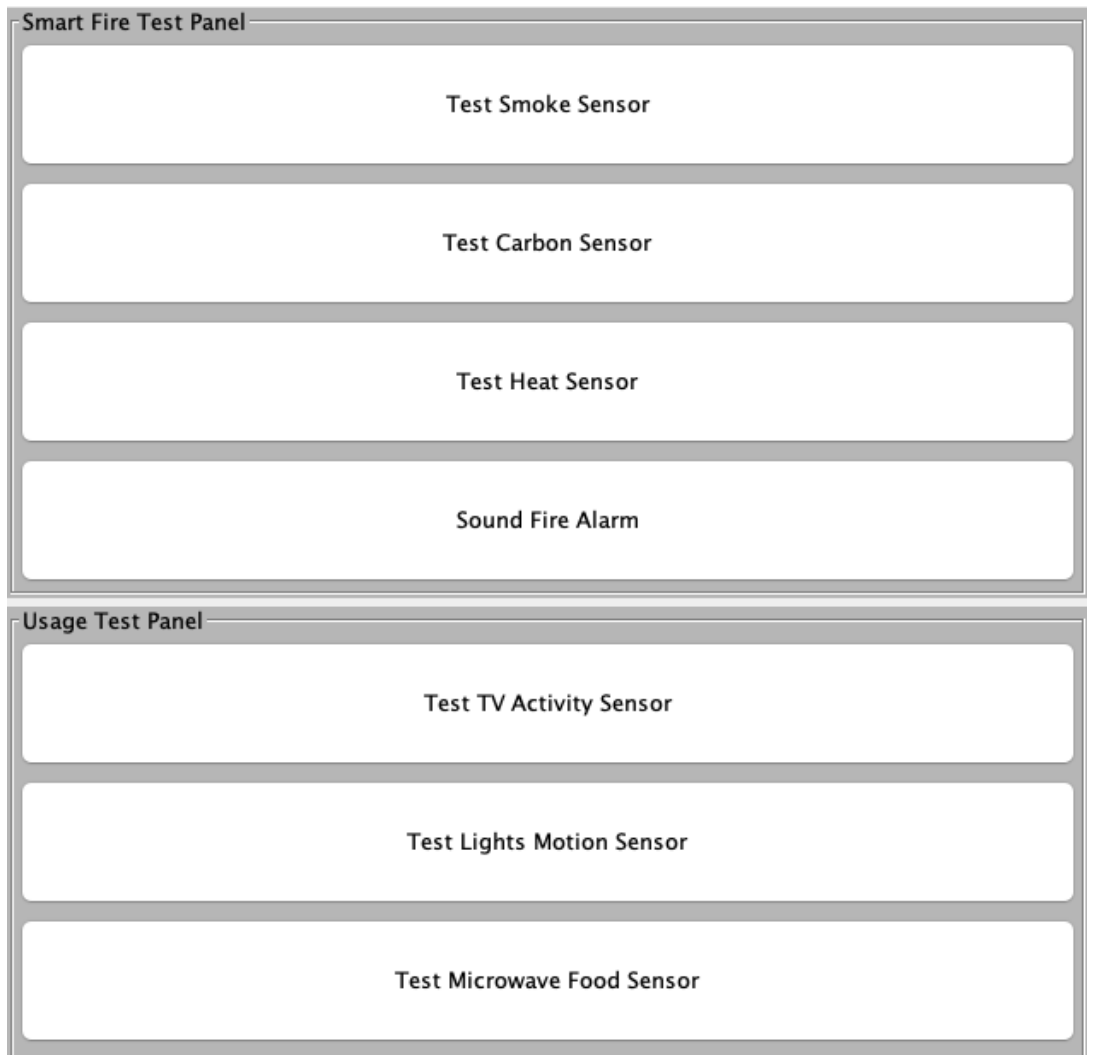

**Figure 20.** Smart home system: simulation dashboard with sample testing panels.

Song et al. (31) contribute in the IoT domain, particularly within the domain of smart homes. This work focuses on the rapid development of IoT in smart homes by discussing family networking technology and proposing typical smart home system solutions. Additionally, they have conducted a modelling and simulation study for WiFi and LTE coexistence scenarios, however, the study is still in the exploratory stage with many solutions yet to be validated.

Rouillard et al. (32) conducted a smart home simulation study using SCXML statechart diagrams. However, their primary focus was on the communication between the user and smart systems, specifically for sending commands. They presented an overall statechart for simulating user interaction with the smart system, using three entities: action, object, and place. This statechart outlines the actions to be performed on an object at a specific location.

Corno et al. (33) addresses the increasing complexity of Smart Environments (SmE) and emphasizes the significance of correctness, reliability, safety, and security in SmE applications. It proposes a design-time modelling and formal verification methodology to ensure error-free and requirement-compliant implementation of SmE systems. This methodology utilizes various modelling approaches, such as ontology and statecharts, and employs model checking to verify SmE components against specified requirements.

While the above approaches provide modelling, analysis, or simulation methods for IoT systems with a focus on smart home application, unlike STL4IoT, they do not facilitate the system design activity with templates or patterns customized for the target domain.

### *Summary*

A comparison of the related work based on several criteria is presented in Table 1. The review considered support for and relevance to the following: system design with the use of modelling templates or model composition techniques; IoT and CPS domains; modelling, simulation, and code generation; and case studies or applications used for validation. Note that *SC* in the table refers to statecharts.

There is limited work available for facilitating the design process of such complex systems using model-based simulation techniques. Some existing research work have adopted statecharts to design the intricate dynamics of IoT and CPS systems. Their focus extend to modelling user behaviour, environmental influences, control mechanisms, and interactions, underscoring the pivotal role of these elements in fortifying system models. Several of these works prioritize code generation, bridging the conceptual models with practical implementations. There is indeed a need for templates or patterns that support the design process along with a comprehensive framework for modelling, simulating, and generating code in the IoT and CPS domains. STL4IoT offers an extensive template library and toolkit for modelling and simulation of IoT systems. By leveraging statecharts and innovative modelling strategies, this research aims to make a meaningful contribution to the continuously evolving fields of IoT and CPS, providing a versatile tool for modelling a diverse array of smart systems.

## Conclusion

The paper proposes a library of statechart templates to model the behaviour of cyber-physical and IoT systems. Two templates, *IoT template* and *Hub template*, have been developed that provide skeletons for modelling statecharts for smart systems as well as smart hubs (e.g, a smart hub coordinating multiple smart systems in a home). An extensible library of basic atomic components modelling the different aspects of IoT systems, including hardware, software, and network, has been proposed. A generic statechart template that is reusable for various IoT systems has been designed for each of the atomic component, namely, sensor, actuator, controller, physical entity, network, and power calculator. We demonstrate the use of our templates with a representative case of a smart home system composed of multiple smart systems (smart lights, smart TV, smart microwave, and smart fire alarm) including a smart hub of lights. The template can be used to design systems outside the smart home domain.

As a next step, we intend to extend the set of templates to cover systems with a broader range of components, hardware devices (e.g., tags) and network connectivity (e.g., LPWAN and Zigbee). For further evaluations, we plan on using the templates to design training simulators for more complex systems, such as a smart city. Moreover as future work, we will be modelling the architectural view of the system along with the statechart model to go towards synthesizing code for deployment based on the architecture.

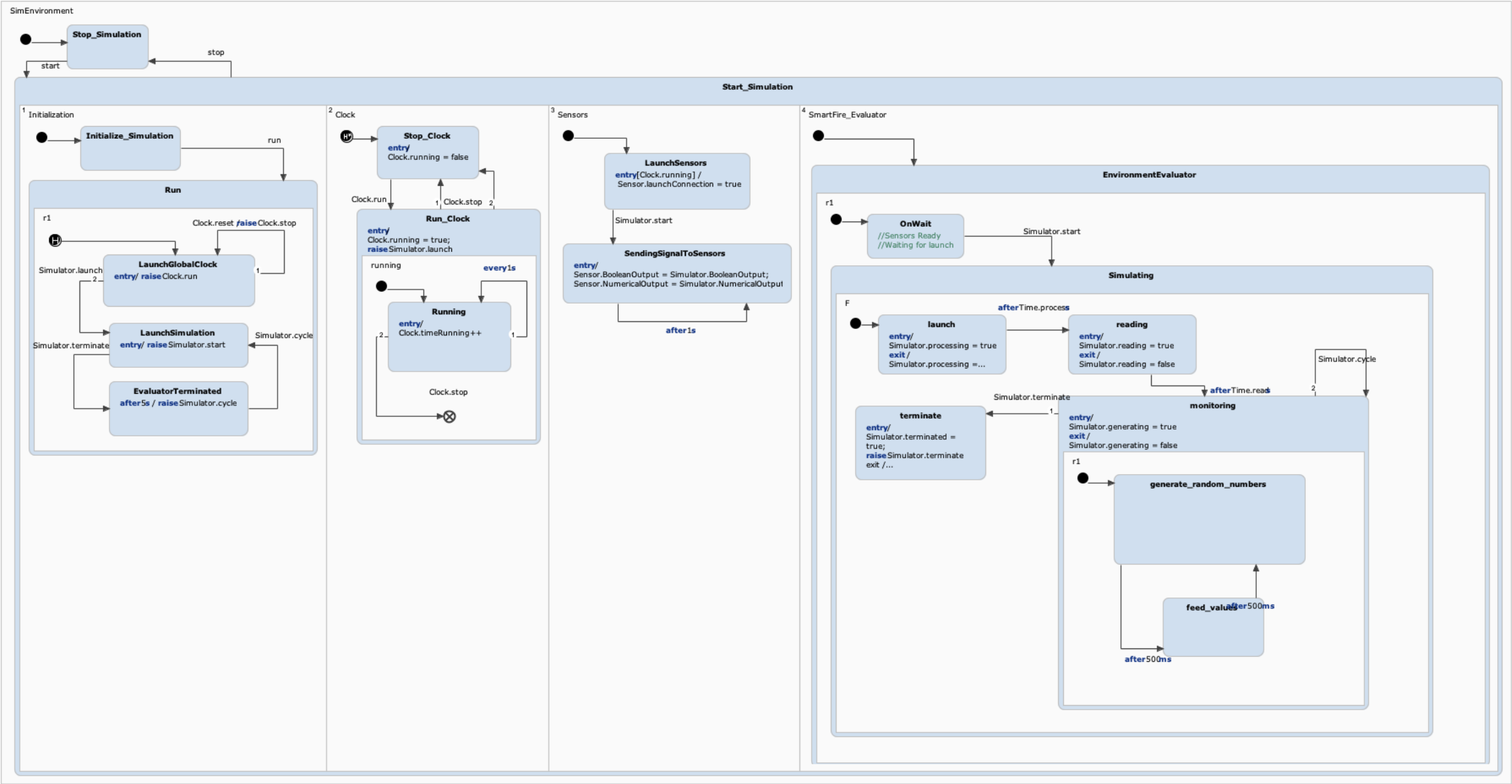

**Figure 21.** Simulation environment: statechart model.

**Table 1.** Comparison of the approaches (supports (✓), does not support (×), unknown/unclear (-)).

| Approach | Templates | | Composition | | Domain | | M&S | | Code Gen | Case Study |
|---|---|---|---|---|---|---|---|---|---|---|
| | SC | Other | SC | Other | IoT | CPS | Modelling | Simulation | | |
| (Gamma) Molnar et al. (18), Graics et al. (19) | × | × | × | ✓(GCD) | × | × | ✓(SC) | × | ✓ | Traffic Light System |
| Csuvarski et al. (20) | × | × | × | ✓(Gamma) | × | ✓ | ✓(SC-based) | ✓ | ✓ | Crossroad System |
| Dinkelbach et al. (21) | × | ✓ | × | × | - | ✓ | ✓(CIM) | × | ✓ | Smart Power Grid System |
| (OpenMETA) Sztipanovits et al. (23) | × | × | × | ✓ | × | ✓ | ✓(CyPhyML) | ✓ | × | *VehicleForge* |
| (IoTMOF) Boutot (25) | × | × | × | × | ✓ | ✓ | ✓(UCM4IoT) | ✓ | ✓ | Smart Light System |
| Xiao et al. (26) | × | × | × | × | ✓ | ✓ | ✓(FSM) | ✓ | × | Smart Blind, LED Actuator, Light Sensor |
| Almeida et al. (34),(27) | × | × | × | × | ✓ | × | ✓(FSM) | ✓ | × | Multimodal Devices |
| Sinha et al. (28) | × | × | ✓ | × | ✓ | × | ✓(SC) | ✓ | × | Fall Detection System |
| Costa et al. (29) | × | × | × | × | ✓ | × | ✓(SysML4IoT) | × | ✓ | Energy Conservation Application |
| Harrand et al. (30) | × | × | × | × | ✓ | × | ✓(ThingML) | × | ✓ | E-Health and Home Automation System |
| Song et al. (31) | × | × | × | × | ✓ | × | ✓ | ✓ | × | Smart Home System |
| Rouillard et al. (32) | × | × | ✓ | × | ✓ | × | ✓ | ✓ | ✓ | Smart Digital Home |
| Corno et al. (33) | × | × | × | × | ✓ | × | ✓(SCXML) | × | × | Smart Environment |
| (xMAS) Chatterjee et al. (22) | × | × | × | ✓ | × | × | ✓ | ✓ | × | *Ring, SMF, IOF, NC, TD* |
| (RSARTE) Mohlin et al. (24) | × | ✓ | × | ✓ | - | - | × | × | ✓ | - |
| (STL4IoT) Rempillo and Mustafiz (6) | ✓ | × | ✓ | × | ✓ | ✓ | ✓(SC) | ✓ | ✓ | Smart Home System |

## Copyright

### Acknowledgements

This work is partly funded by NSERC and Toronto Metropolitan University. The authors would like to thank Prof. Hans Vangheluwe for his valuable insights on this work.